\documentclass[twocolumn]{aastex701}

\def\chandra    {{\em Chandra}\/}

\def\xmm        {XMM-{\em Newton}\/}

\def\arcdeg{\hbox{$^\circ$}}
\def\arcmin{\hbox{$^\prime$}}
\def\arcsec{\hbox{$^{\prime\prime}$}}

\def\arcdeg{\hbox{$^\circ$}}
\def\arcmin{\hbox{$^\prime$}}
\def\arcsec{\hbox{$^{\prime\prime}$}}

\begin{document}

\title{AGN feedback destroying group cool cores --- a case study on IC~2476 with a remnant radio AGN and an exceptionally low abundance}

\author[orcid=0000-0002-0924-9668,sname='Liu']{Wenhao Liu}
\affiliation{Purple Mountain Observatory, Chinese Academy of Sciences, Nanjing, Jiangsu 210023, People's Republic of China}
\affiliation{School of Astronomy and Space Sciences, University of Science and Technology of China, Hefei 230026, People's Republic of China}
\email[show]{whliu@pmo.ac.cn}  

\author[orcid=0000-0001-5880-0703,sname='Sun']{Ming Sun} 
\affiliation{Department of Physics and Astronomy, University of Alabama in Huntsville, Huntsville, AL 35899, USA}
\email[show]{ms0071@uah.edu}

\author[orcid=0000-0003-0297-4493,sname='Nulsen']{Paul Nulsen}
\affiliation{ICRAR, University of Western Australia, 35 Stirling Hwy, Crawley, WA 6009, Australia}
\affiliation{Center for Astrophysics | Harvard and Smithsonian, 60 Garden Street, Cambridge, MA 02138, USA}
\email{paulnulsen@gmail.com}

\author[orcid=0000-0003-4120-9970,sname='Brienza']{Marisa Brienza}
\affiliation{INAF - Istituto di Radioastronomia, via P. Gobetti 101, 40129, Bologna, Italy}
\email{marisa.brienza@inaf.it}

\author[orcid=0000-0003-0167-0981,sname=Sarazin]{Craig Sarazin}
\affiliation{Department of Astronomy, University of Virginia, Charlottesville, VA 22904, USA}
\email{cls7i@virginia.edu}

\author[orcid=0000-0001-7917-3892,sname=Eckert]{Dominique Eckert}
\affiliation{Department of Astronomy, University of Geneva, Ch. d’Ecogia 16, 1290 Versoix, Switzerland}
\email{Dominique.Eckert@unige.ch}

\author[orcid=0000-0001-8176-7665,sname=Tamhane]{Prathamesh Tamhane}
\affiliation{Department of Physics and Astronomy, University of Alabama in Huntsville, Huntsville, AL 35899, USA}
\email{pdt0003@uah.edu}

\author[orcid=0000-0003-3728-9912,sname=Zheng]{Xianzhong Zheng}
\affiliation{Tsung-Dao Lee Institute and Key Laboratory for Particle Physics, Astrophysics and Cosmology, Ministry of Education, 
Shanghai Jiao Tong University, Shanghai 201210, People's Republic of China}
\email{xzzheng@sjtu.edu.cn}


\begin{abstract}
We present results from deep \xmm\ observations ($\sim180$ ksec) of the galaxy group IC~2476 in the X-GAP sample, 
capturing a rare AGN feedback regime, where the central radio AGN has ceased activity but the radio lobes are still 
detectable. The hot intragroup medium of IC~2476 features a compact cool core, with a size of $\sim 10$ kpc, 
and a moderate X-ray luminosity of $\sim3.4\times10^{40}$ erg~s$^{-1}$. We measured an unusually low temperature
of $\sim$0.45 keV within the cool core. We measured an exceptionally low abundance of below $\sim0.2$ solar, even in the 
central region. The X-ray surface brightness of the group is not symmetric and shows 
a clear depression in the region overlaid with the eastern radio lobe, suggesting the presence of an X-ray cavity. 
The scaled entropy of the hot gas within 0.1 $r_{500}$ of IC~2476 is among the highest in X-ray galaxy groups with a 
similar system temperature. Our study suggests that the past powerful radio AGN outburst in IC~2476 has disrupted the cool core
and the X-ray cavity has dissipated or mixed with the surrounding hot gas since the shutdown of the central AGN. 
A large cool core may be gradually re-established in the absence of AGN activity. This system therefore provides a great
example of AGN feedback destroying the group cool core, even with a moderate radio luminosity at present.
\end{abstract}



\section{Introduction} 

\noindent
Feedback from the central active galactic nuclei (AGN) has been widely accepted to play an important role in galaxy formation
and evolution \citep[e.g.,][]{McNamara07,Fabian12,Gaspari2020,Eckert2021,Hlavacek2022, Donahue2022,Heckman2023,Harrison2024}.
AGN feedback may simultaneously explain the quenching of star formation in massive galaxies, the exponential cut-off
at the bright end of the galaxy luminosity function, the supermassive black hole mass-bulge mass relation, and the quenching of 
cooling-flows in cluster cores \citep[e.g.,][]{Scannapieco05,Begelman05,Croton06}.
The radio jets launched from the central brightest galaxies of clusters/groups can inflate radio lobes and create disturbances, 
such as shocks and cavities in the intracluster/group medium, which can be observed through X-rays. These features
serve as calorimeters for the total energy output of AGN that allow us to understand a great deal of AGN
feedback and SMBH growth \citep[e.g.,][]{McNamara07}.
In the nearby Universe, almost all cool core clusters host a radio-loud AGN.

However, in contrast to the numerous detailed analysis of clusters, AGN feedback has not been well studied in galaxy groups. Galaxy
groups are not simply scaled-down versions of clusters and AGN feedback might operate differently in them. Due to the shallower
gravitational potential of galaxy groups, the impact of AGN outbursts can be much more pronounced in groups
\citep[e.g.,][]{Giodini2010}.
However, direct evidence of AGN heating in groups is generally limited to bright X-ray
cool cores with weak radio AGN. 
Galaxy groups are gas poor around the center and group cool cores are smaller than cluster cool cores. \citet{Sun2009}
showed that groups with a luminous cool core do not host a luminous radio AGN. The groups with luminous radio AGN only have small 
coronae or small cool cores, so their extended X-ray atmospheres that enclose the radio lobes are faint. Both cavities and
shocks are more easily detected in high density regions, so these features become subtle in groups, especially
for groups hosting strong radio AGN, making them much more difficult to study.

Based on a \chandra\ study of a large nearby cluster sample, \citet{Liu2024} proposed a transition from large X-ray cool cores to cool 
core coronae due to radio AGN heating in galaxy clusters. 
Strong radio AGN can deposit a large amount of energy outside the central region and, thus, can disrupt the large X-ray cool cores of a galaxy cluster, 
leaving cool core remnants. Due to their shallower gravitational potentials, group cool cores
could be more easily destroyed by an outburst from a radio AGN. In addition, the central cooling times of galaxy groups are found to be less than 1 Gyr 
in most cases \citep[e.g.,][]{Sun12}, suggesting that the central AGN in galaxy groups can launch radio jets more often than in galaxy clusters. 
Therefore, the duty cycle, defined as the fraction of the time when the radio AGN is active, is likely different in galaxy groups.
The group X-ray cool cores could be disrupted and turned into a cool core remnant more often during the radio AGN feedback cycle.

IC 2476 (also B2 0924+30) is one of the first identified remnant radio galaxies \citep{Cordey1987} and is hosted by the BCG of a galaxy group. 
It has a curved radio spectrum.
Between 610 MHz and 1415 MHz, the spectral index $\alpha$ (defined as $S\propto \nu^{-\alpha}$ where
$S$ is the radio flux and $\nu$ is the frequency) varies from 1.2 in the brightest parts of the lobes to
1.7 near the center of the source. Follow-up VLA observations from 1.5 GHz to 8.5 GHz failed to detect
any hotspots, radio jets or the radio nucleus \citep{Cordey1987}. Thus, the central engine of IC~2476
has shut down but the radio lobes have not faded away.
Based on the spectral index distribution and spectral ageing maps, \citet{Shulevski2017} classified this source as an FR-II AGN remnant and 
showed that the AGN activity ceased around 50 Myr ago \citep[see also][]{Jamrozy2004}.
Using a new approach that combines spectral ageing and dynamical models, \citet{Turner2018} showed that 
IC 2476 has an active phase lasting about 50 Myr and has been in a quiescent phase for about 28 Myr.
Therefore, IC~2476 is at a particular stage of AGN feedback cycle and it is an excellent target for us to study the cycle of radio AGN feedback.

In this paper, we use deep \xmm\ observations to study AGN feedback in this remnant radio system.
The galaxy group IC~2476 has a low temperature ($\sim1$ keV). Based on the $L-T$ relation \citep{Lovisari2015}, its 0.1-2.4 keV luminosity 
is $3.5\times10^{42}$ erg s$^{-1}$. The general properties of IC~2476 are summarized in Table \ref{tab1}. 
The paper is organized as follows. The observations and data preparation are presented in Section \ref{Observation}. 
Data analysis is described in Section \ref{Analysis}.
In Section \ref{Results} we present the results.
The Discussion is in Section \ref{discussion} and the Conclusions in Section \ref{Conclusion}. 
Throughout the paper, we assume a cosmology with $H_{0}=70$ km s$^{-1}$ Mpc$^{-1}$, $\Omega_{M}=0.3$, and $\Omega_{\Lambda}=0.7$.
At a redshift of $z=0.0266$, the luminosity distance is 116.248 Mpc and 1\arcsec\ corresponds to 0.535 kpc. All error bars are quoted at the 1$\sigma$
confidence level, unless otherwise specified.
\begin{figure}
\begin{center}
\hbox{\hspace{-2px}
\includegraphics[scale=0.30]{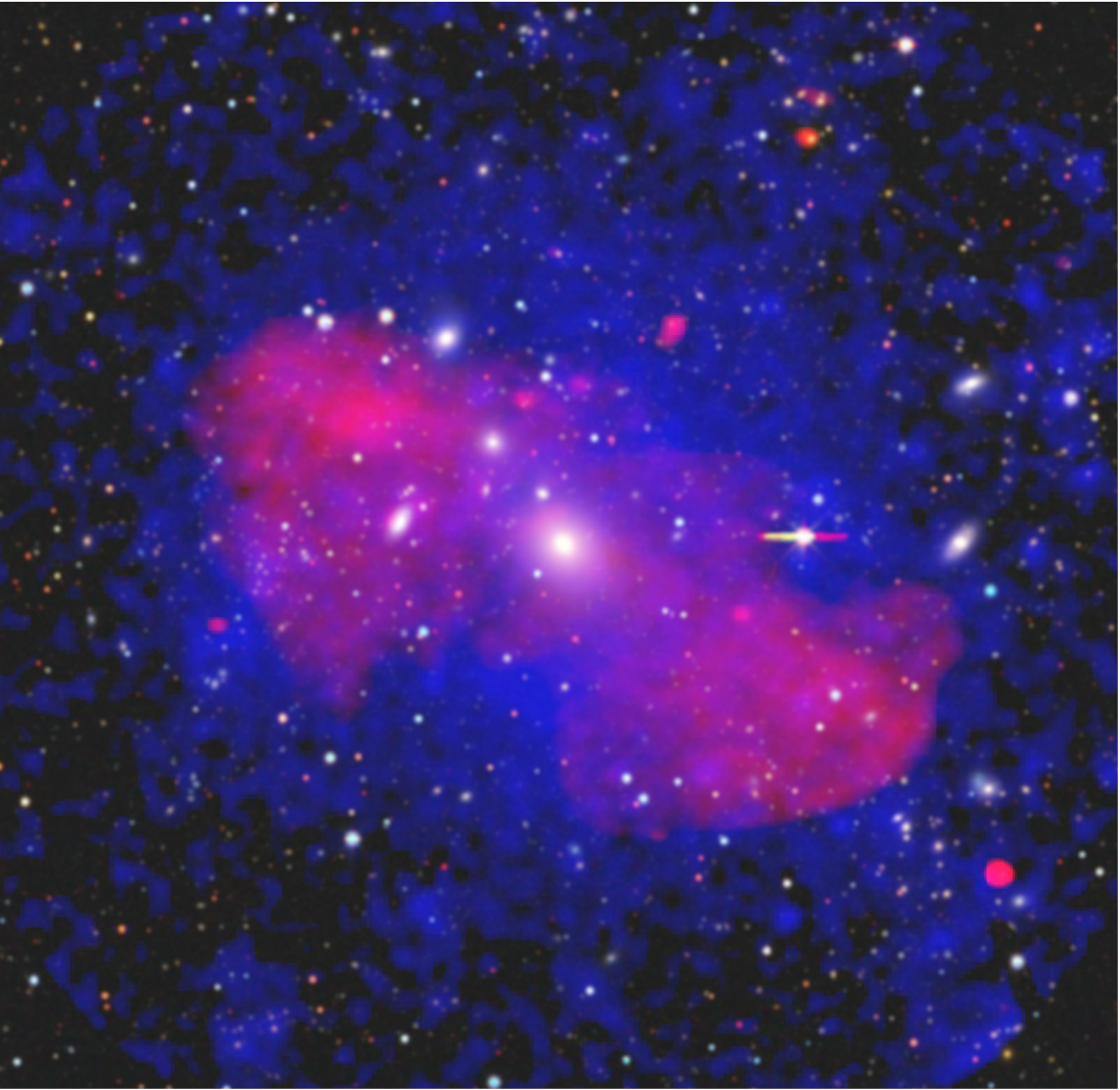}
}
\caption{Three-color composite radio/optical/X-ray image of galaxy group IC~2476.
	Red: LOFAR radio image at 144 MHz. White: combined optical image from DESI legacy imaging surveys. 
	Blue: \xmm\ 0.4–3.0 keV image smoothed with a two-dimensional Gaussian 5-pixel kernel. The removed point sources have been filled with the background for visual purposes.
}
\label{fig1}
\end{center}
\end{figure}


\begin{deluxetable*}{ccccccc}
\tablecaption{Properties of Galaxy Group IC~2476}
\tablehead{
     \colhead{$z$} & \colhead{$\sigma$\tablenotemark{a}}  & \colhead{$P_{1.4GHz}$\tablenotemark{b}} &  \colhead{$M_{\star}$\tablenotemark{c}} & \colhead{$M_{500}$\tablenotemark{d}} & \colhead{$kT$\tablenotemark{e}} & \colhead{$L_{\rm 0.5-2 keV}$\tablenotemark{f}} \\
	\colhead{}       & \colhead{(km/s)} & \colhead{(W Hz$^{-1}$)} & \colhead{($10^{11}$ M$_\odot$)} & \colhead{($10^{13}$ M$_\odot$)} & \colhead{(keV)} & \colhead{($10^{42}$ erg s$^{-1}$)}
}
\startdata
	 0.0266 & $260.8\pm42.2$ & 6.9E23, $<$6.7E20 & 2.3 & 2.1 & $1.00_{-0.01}^{+0.02}$, $1.03_{-0.03}^{+0.02}$  &  $2.25_{-0.08}^{+0.14}$, $2.71_{-0.18}^{+0.12}$ \\
\enddata
	\tablenotetext{a}{Velocity dispersion measured from 30 galaxies within 2$r_{500}$ estimated from the NASA/IPAC Extragalactic Database (NED); }
    \tablenotetext{b}{The total and core radio power from \citet{Capetti2002};}
    \tablenotetext{c}{The stellar mass of the central galaxy from the galaxy group catalog by \citet{Tinker2021};}
	\tablenotetext{d}{Estimated from $M-T$ relation \citep{Sun09};}
	\tablenotetext{e}{System temperature measured within $0.15-1.0$ $r_{500}$ and within $r_{500}$;}
	\tablenotetext{f}{X-ray luminosity in the rest-frame $0.5-2$ keV band measured within $0.15-1.0$ $r_{500}$ and within $r_{500}$; The bolometric luminosity is 2.35 times larger;}
\label{tab1}
	\vspace{-20px}
\end{deluxetable*}

\section{Observations and Data Preparation}
\label{Observation}
IC 2476 was observed by \xmm\ for $\sim90$ ksec in October 2020 and for $\sim90$ ksec in May 2021 (Obs. IDs 0862260201, 0862260301, PI: W. Liu). 
There was an \xmm\ observation taken in 2009 with an exposure of $\sim50$ ksec, which is also included in our study. The details of the \xmm\ 
observations used in this study are summarized in Table \ref{tab_obs}.
\begin{deluxetable}{cccc}
\tablecaption{Summary of \xmm\ Observations of Galaxy Group IC~2476}
\tablehead{
\colhead{ObsID} & \colhead{Date} & \colhead{Exp}    & \colhead{Cleaned Exp\tablenotemark{a}} \\
	\colhead{}      & \colhead{}     & \colhead{(ksec)}         & \colhead{(ksec)}        
}
\startdata
0862260201 & 2020 Oct 21 & 88.9 & 76.6, 78.9, 66.3 \\
0862260301 & 2021 May 04 & 87.7 & 76.4, 78.9, 66.6 \\
0604530101 & 2009 Nov 26 & 47.5 & 26.9, 27.2, 22.4 \\
\enddata
\tablenotetext{a}{the cleaned exposure time of MOS1, MOS2, and PN.}
\label{tab_obs}
\end{deluxetable}

The \xmm\ data was processed using the Extended Source Analysis Software \citep[ESAS;][]{Snowden2008, Kuntz2008}, 
which is integrated into the \xmm\ Science Analysis System (SAS; version 19.0.0). 
The raw event files from the MOS and PN CCDs were re-processed using tasks {\tt emchain} and {\tt epchain}, with the latest calibration made available in April 2021 applied.
The solar soft proton flares were filtered out with {\tt mos-filter} and {\tt pn-filter}, respectively. Point sources were detected using the task {\tt cheese}, then visually 
inspected and excluded. The count images and exposure maps were produced by {\tt mos-spectra} and {\tt pn-spectra},
which were also used to extract the spectra and response files. The instrumental background images and spectra were modeled with the tasks {\tt mos\_back}
and {\tt pn\_back}. The count images, background images, and exposure maps from the MOS and PN, from multiple observations, were combined with the tool
{\tt merge\_comp\_xmm}. The final background-subtracted, exposure-corrected, and smoothed images were generated using the tool {\tt adapt}.
The spectra extracted from the MOS and PN were fitted jointly with the {\tt XSPEC} \citep{xspec} package (version 12.11.1). The database of atomic data ATOMDB version 3.0.9 
and the solar abundance table from \citet{aspl} were used.

In order to estimate the contribution of the cosmic X-ray background, we extracted spectra from an annulus 12\arcmin-14\arcmin\ from the BCG for the MOS and PN detectors, 
where the emission from the galaxy groups is negligible. 
The RASS background spectrum was extracted from a region located between 1 and 1.5 degrees from IC~2476, using the 
X-ray background tool\footnote{https://heasarc.gsfc.nasa.gov/cgi-bin/Tools/xraybg/xraybg.pl}. 
We fitted the EPIC and RASS spectra jointly with background models including an unabsorbed APEC \citep{Smith2001} model with a temperature of 0.1 keV for
the local hot bubble \citep[e.g.,][]{Liu2017}, an absorbed APEC model with a fixed temperature of 0.2 keV for the Galactic halo, 
and an absorbed power-law with a fixed photon index of 1.46 for the cosmic X-ray background. 
The absorption model is {\tt TBABS} with a fixed column density of $1.89\times10^{20}$ cm$^{-2}$, determined using the {\tt NHtot} tool\footnote{http://www.swift.ac.uk/analysis/nhtot/index.php}, which allows for both the atomic hydrogen and the molecular hydrogen column densities \citep{Willingale2013}.

\section{Spatial and Spectral Analysis}
\label{Analysis}
Fig. \ref{fig1} shows a three-color image of IC~2476, with the background-subtracted, exposure-corrected \xmm\ image in the 0.4-3.0 keV band in blue, an optical 
image obtained from the DESI legacy imaging surveys\footnote{https://www.legacysurvey.org/} in white, 
and the radio image at 144 MHz obtained from the LOFAR Two-metre Sky Survey Second Data Release \citep{Shimwell2022} in red.  The BCG is the bright galaxy at the center of the field.
\begin{figure*}
\begin{center}
\begin{tabular}{ccc}
\includegraphics[width=2.in]{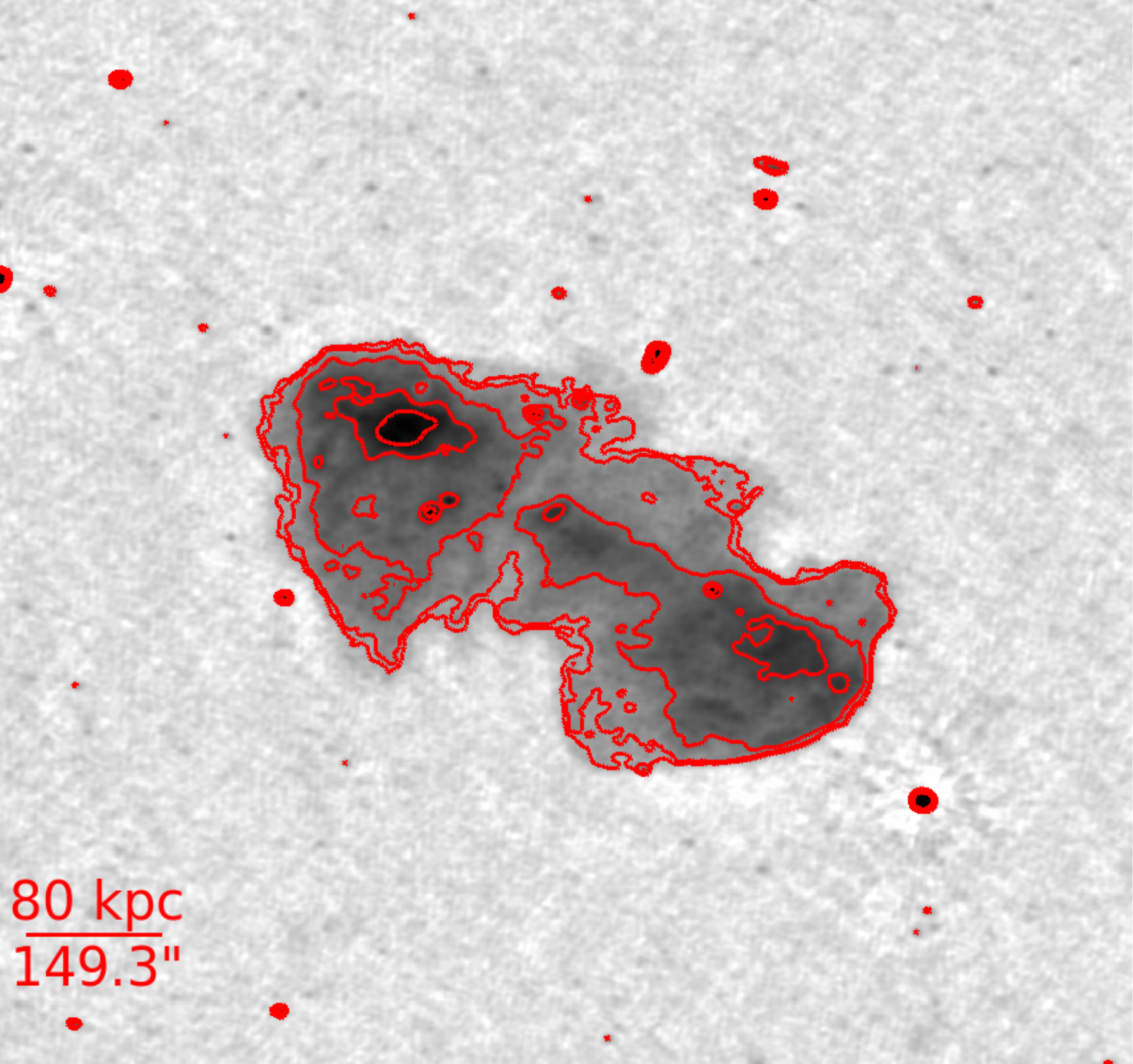} & 
\includegraphics[width=2.3in]{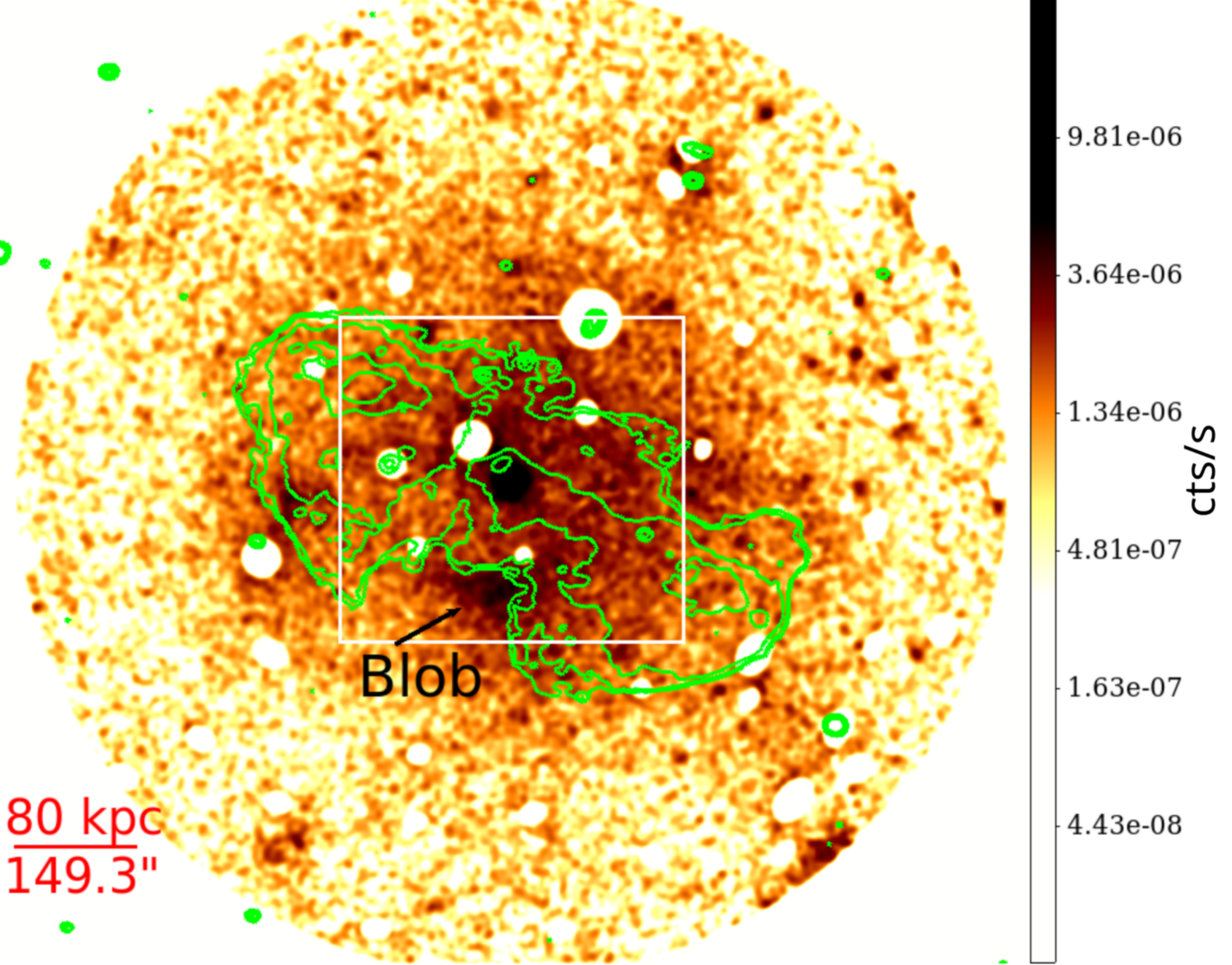}  &
\includegraphics[width=2.3in]{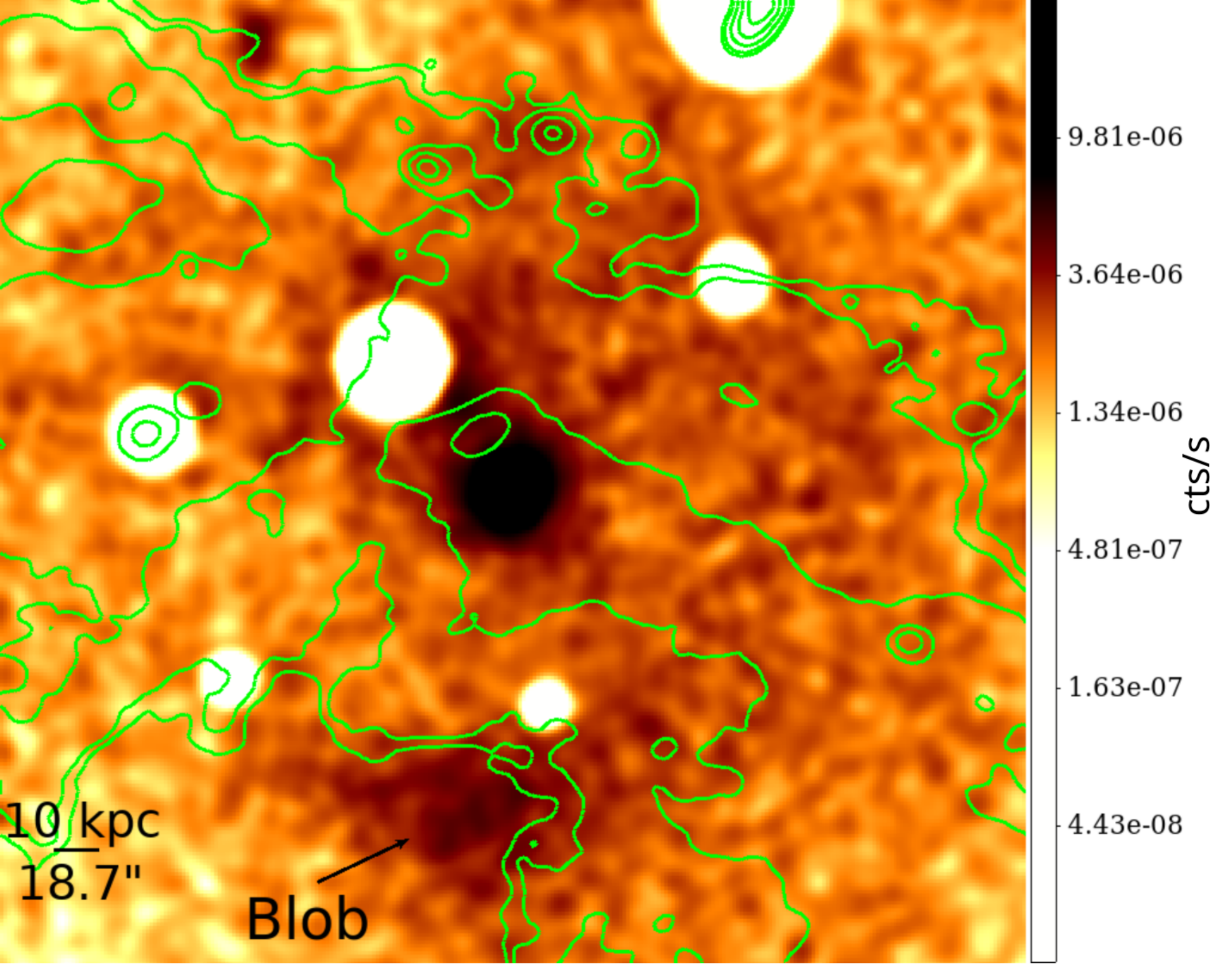} 
\end{tabular}
	\caption{Left: LOFAR image at 144 MHz, with a beam size of 8.6\arcsec$\times$5.5\arcsec, and an rms level (1 $\sigma$) of
	0.1 mJy/beam. Contour levels start from 0.5 mJy/beam to 6 mJy/beam and the square-root spacing is adopted. Middle: 
	The combined background-subtracted, exposure-corrected \xmm\ image of IC~2476 in the 0.4-1.3 keV band, smoothed with a two-dimensional Gaussian 
	3-pixel (2.5\arcsec/pixel) kernel and overlaid with the LOFAR radio contours. There is a blob with enhanced X-ray emission to the south of the central BCG, just outside
	the LOFAR contour. White circular or elliptical regions are X-ray point sources masked (same for other similar figures in this work). The white box marks the region shown in the zoomed-in image in the right panel.
    Right: The zoomed-in \xmm\ image overlaid with the LOFAR
	radio contour, showing the small central cool core, and the blob
	with enhanced X-ray emission to the south of the BCG.
\label{image}}
\end{center}
\end{figure*}

Fig. \ref{image} (left) shows the LOFAR image at 144 MHz of IC~2476. The radio lobes are extending to the northeast and southwest. There are no obvious radio jets, nor radio core.
Fig. \ref{image} (middle) shows the combined background-subtracted, exposure-corrected \xmm\ image of IC~2476, with a zoomed-in image
in the right panel to show the X-ray emission of the central region.
The X-ray image clearly shows that there is a small cool core in the central region with a radius of $\sim10$ kpc. 
There is a blob of enhanced X-ray emission located $\sim80$ kpc to the south
of the central BCG, with an extent of $\sim40$ kpc, which lies just outside the outermost LOFAR radio contour.

\begin{figure*}
\begin{center}
\begin{tabular}{cc}
\hspace{-40px}
\includegraphics[width=2.5in]{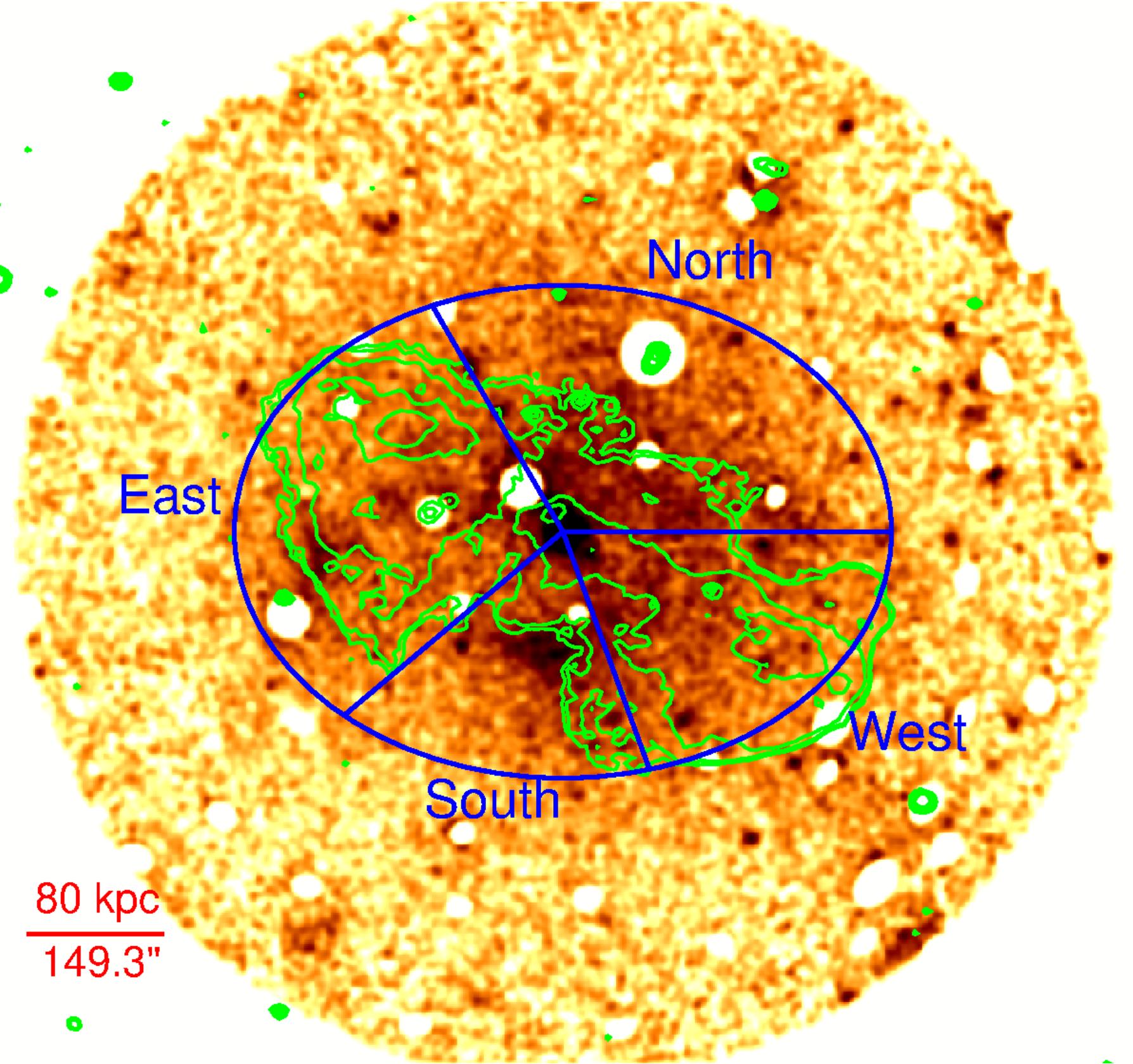} & 
\hspace{50px}
\includegraphics[width=3.3in]{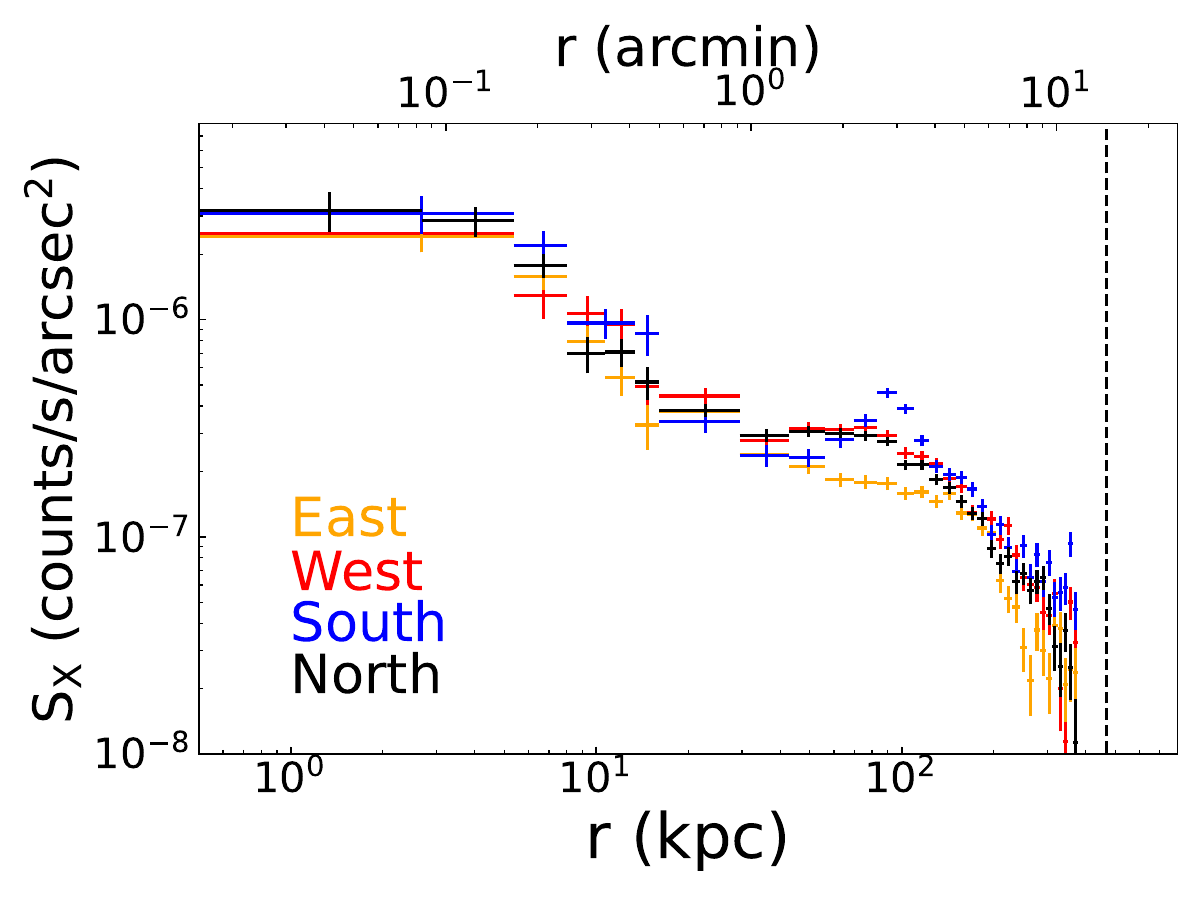} 
\end{tabular}
	\caption{Left: The combined background-subtracted, exposure-corrected \xmm\ image of IC~2476 in the 0.7-1.3 keV band, 
	smoothed with a two-dimensional Gaussian 3-pixel kernel and overlaid with the LOFAR radio contours. The surface brightness profiles
	are extracted in the same ellipse region from four wedges. The ellipse has a major-to-minor axis ratio of 4/3.
	Right: The surface brightness profiles in the 0.7-1.3 keV band from four sectors shown in the left panel. The X-axis is
	the semi-major axis of the ellipse. The dashed line marks $r_{500}$, which is estimated from the scaling relation for 
	galaxy group \citep{Sun09}.
\label{wedge}}
\end{center}
\end{figure*}

\begin{figure}
\vspace{8px}
\begin{center}
\hspace{-10px}
\includegraphics[scale=0.36]{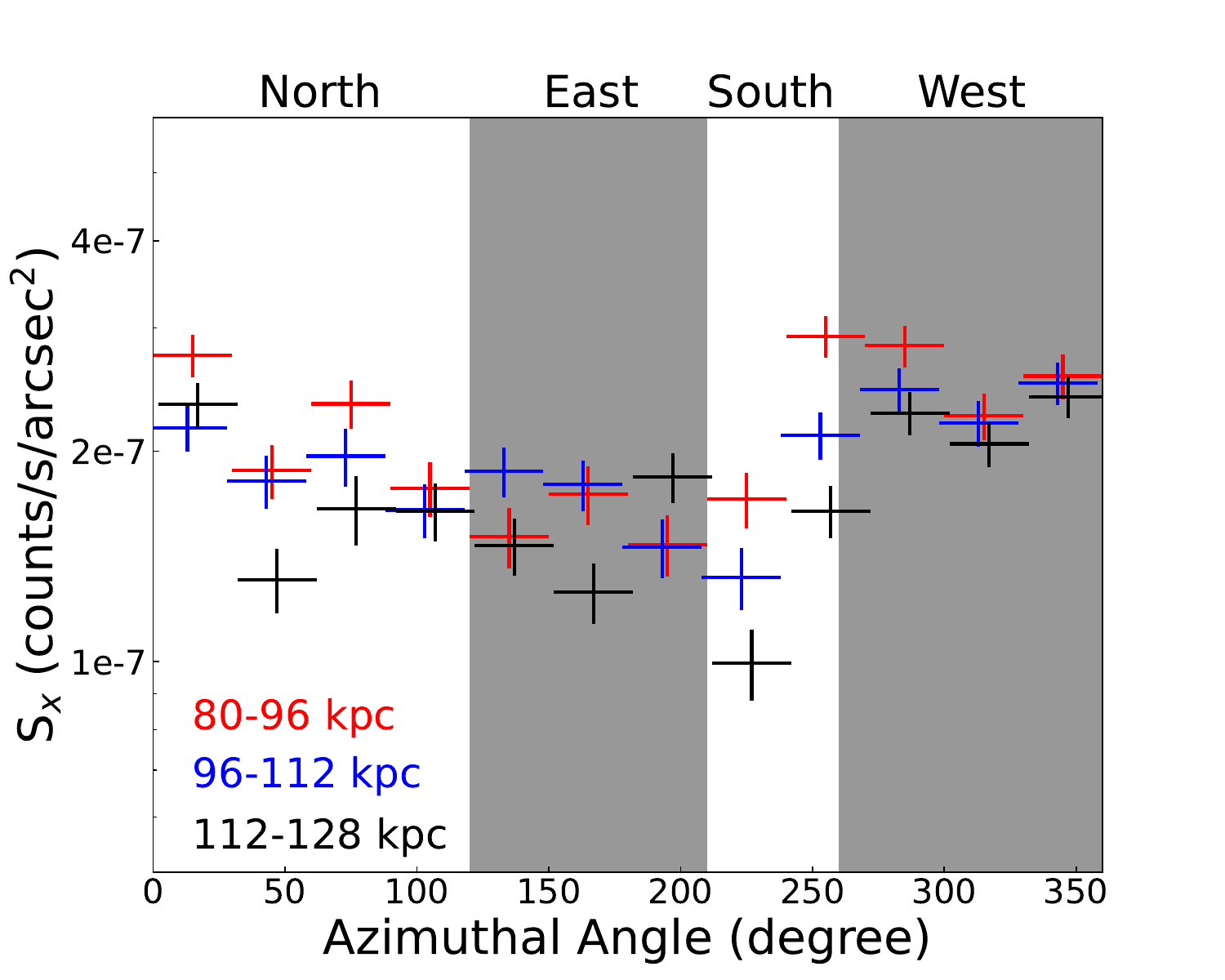} 
	\caption{The surface brightness profiles extracted along the azimuthal directions (measured counterclockwise from the west)
with angle step size of 30 degrees from three annuli with inner and outer radii of (80, 96) kpc (red), (96,112) kpc (blue), and (112, 128) kpc (black) in 0.7-1.3 keV band. The grey regions roughly show the range for the eastern and western radio lobes.
\label{sb_azimuthal}}
\end{center}
\end{figure}

We extracted radial surface brightness profiles in the 0.7-1.3 keV band from the four wedges shown
in Fig. \ref{wedge} (left). The center of the ellipse is chosen to coincide with the optical center of IC~2476.
The eastern and western wedges were chosen to cover the radio lobes. The ratio of the semi-major axis to the semi-minor axis is 4/3, and 
the position angle of the ellipse is 90\arcdeg\ east of north. Fig. \ref{wedge} (right) shows the surface brightness profiles extracted from the four wedges, and we used
the semi-major axis of the ellipse as the X-axis for all profiles. The Non-X-ray particle background has been
subtracted. The cosmic X-ray background is modeled and estimated to have a flux of $\sim5\times10^{-8}$ counts s$^{-1}$ arcsec$^{-2}$,
which has also been subtracted from the surface brightness profiles. We see clear differences between the surface brightness
in the different wedges. The blob mentioned above, with enhanced X-ray emission to the south of the BCG,  appears as a bump
from $\sim80-110$ kpc in the surface brightness from the southern wedge. The surface brightness from the eastern wedge 
is depressed at $60-120$ kpc relative to the profiles from the other
three wedges. In general, the surface brightness is depressed over X-ray cavities, where radio plasma has displaced the hot gas in galaxy clusters and groups.

In Fig. \ref{sb_azimuthal} we show the surface brightness profiles extracted along the azimuthal directions with an angle step size of 30 degrees 
from three annuli from 80-128 kpc, each with a radial size of 16 kpc. However, we see no clear boundaries around the X-ray cavity, 
suggesting that the boundary may have become diffuse as the result of mixing between the radio plasma and the neighboring gas.

\begin{figure*}
\begin{center}
\hbox{\hspace{-5px}
\includegraphics[scale=0.44]{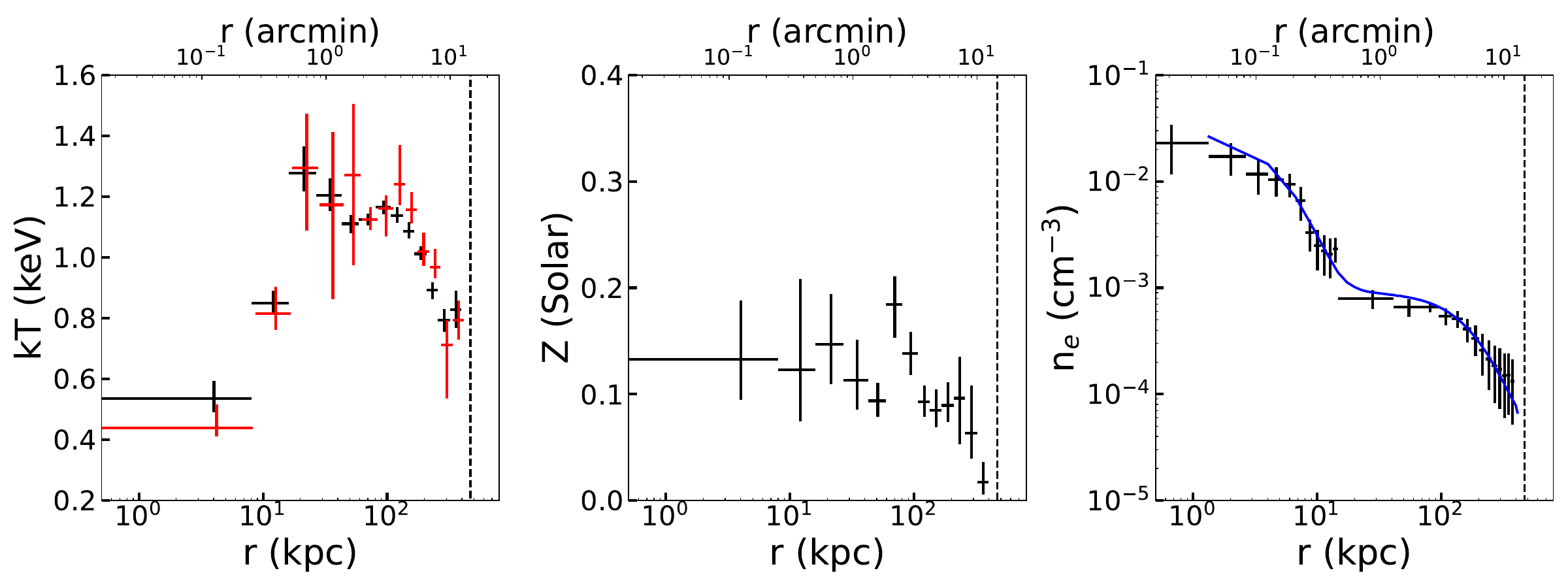}
}
\caption{
	Left: The projected (black) and deprojected (red) temperature profile of IC~2476. We slightly shifted the red points horizontally for 
	display purpose. The dashed line marks $r_{500}$. Middle: The corresponding projected abundance profile.
	Right: The deprojected gas density of IC~2476 from the azimuthally-averaged surface brightness profile. 
	The dashed line marks $r_{500}$. The blue line shows the profile deconvolved
	from the \xmm\ EPIC PSF.
}
\label{kT_ne}
\end{center}
\end{figure*}

To obtain the temperature and gas density profiles for IC~2476, we followed the method used in \citet{Liu2024}.
First, we extracted spectra from a set of 13 annular regions centered on the
BCG out to a radius of $\sim$400 kpc, from the MOS and PN detectors, for each observation. The projected spectra from the same regions for the
MOS and PN detectors from each observation were fitted simultaneously to an absorbed thermal APEC model, i.e.,
TBABS*APEC, plus the fixed cosmic background models scaled to the extraction area. 
The temperature, metallicity, and the normalization were allowed to vary freely. 
Remaining differences between the responses of the MOS and PN detectors were allowed for with multiplicative constant models. Fig.~\ref{kT_ne} (left) 
shows the azimuthally-averaged radial profiles of the projected temperature of IC~2476. 
The temperature in the innermost bin is $0.54^{+0.06}_{-0.05}$ keV. It rises quickly to $1.27$ keV at $\sim21$ kpc.
The temperature then decreases to $\sim1.1$ keV at $\sim50$ kpc, then rises again until $\sim95$ kpc, before decreasing to $\sim0.8$ keV at large radii. 
Fig.~\ref{kT_ne} (middle) shows the corresponding measured abundances. We found the abundances are exceptionally low (below $\sim0.2$ solar) even in the central 
region. We also show the deprojected temperature profile of IC~2476 in the left panel of Fig.~\ref{kT_ne}. For temperature deprojection, we adopted a 
model-independent approach \citep[e.g.,][]{Sun03, Dasadia16,Liu_3C88} which can account for point source masks in different bins. In two innermost bins, 
the deprojected temperatures are lower than the projected ones, while in other region the deprojected temperatures are generally consistent with the projected 
ones except with larger errorbars. The abundances cannot be well constrained in deprojection fitting and we fixed it to 0.1 solar.

Second, we extracted the azimuthally-averaged surface brightness profile in the 0.7-1.3 keV band, using concentric annular
bins centered on the BCG and subtracted the contribution from the cosmic X-ray background.
To correct for the effect of the point spread function (PSF) in the EPIC camera on-board of \xmm, we followed the method used in \citet{Eckert2020}. 
The PSF only affects the surface brightness within the central $\sim6$\arcsec. 
We then deprojected the surface brightness profiles using the ``onion-peeling''
method to derive the gas density profile \citep[e.g.,][]{Kriss1983, Cavagnolo09, Liu2024}.
We computed the energy conversion factor to take into account variations in temperature and metallicity at each radius, which are obtained from spectral fitting. 
The typical metal abundance is 0.1 solar. The deprojected gas density profile
is shown in the right panel of Fig.~\ref{kT_ne} and the PSF-corrected gas density is shown in blue.
The density profile shows two distinct gas components, with a steep high-density component in
the central $\sim15$ kpc and a relatively flat, low-density component beyond the central region.
This form of density profile is typically seen in cool core corona systems \citep[e.g.,][]{Sun2009}.
\begin{figure*}
\begin{center}
\hbox{\hspace{-5px}
	\includegraphics[scale=0.44]{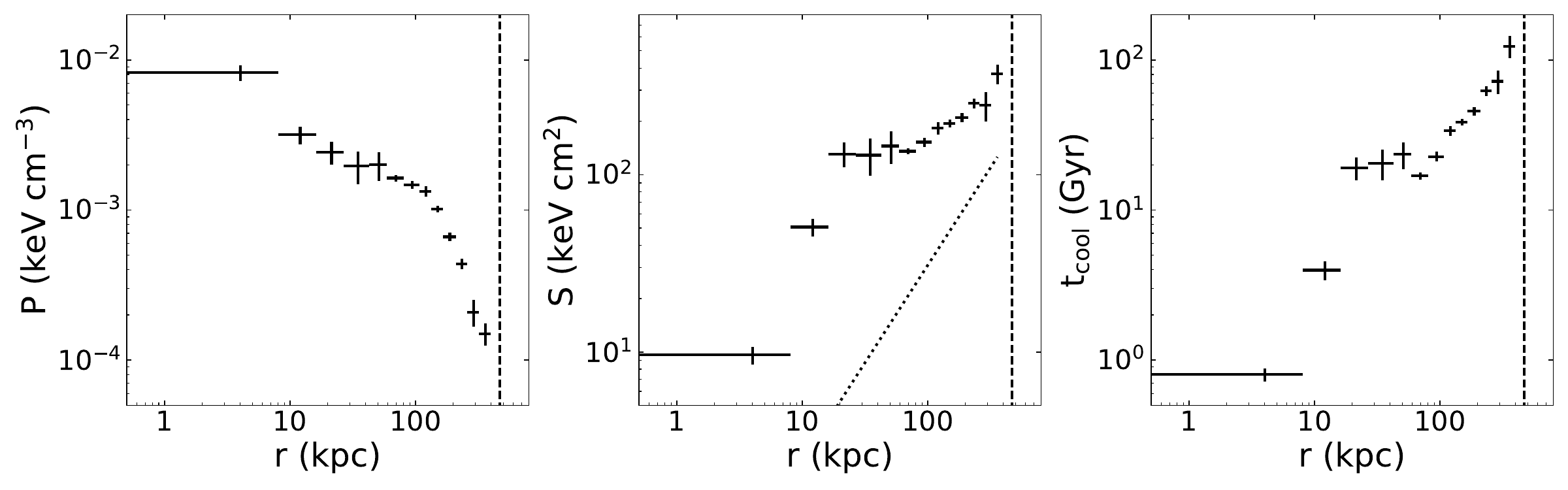}
}
\caption{Radial profiles of pressure (left), entropy (middle), and cooling time (right) of galaxy group IC~2476. 
	The dotted line shows the baseline entropy profile expected from gravitational collapse \citep{Voit05}
	which has been scaled to the group mass (see Equation \ref{eq1}).
	The dashed line marks the radius of $r_{500}$.
}
\label{PST}
\end{center}
\end{figure*}

\begin{figure}
\begin{center}
\vspace{-10px}
\hspace{-30px}
\includegraphics[scale=0.35]{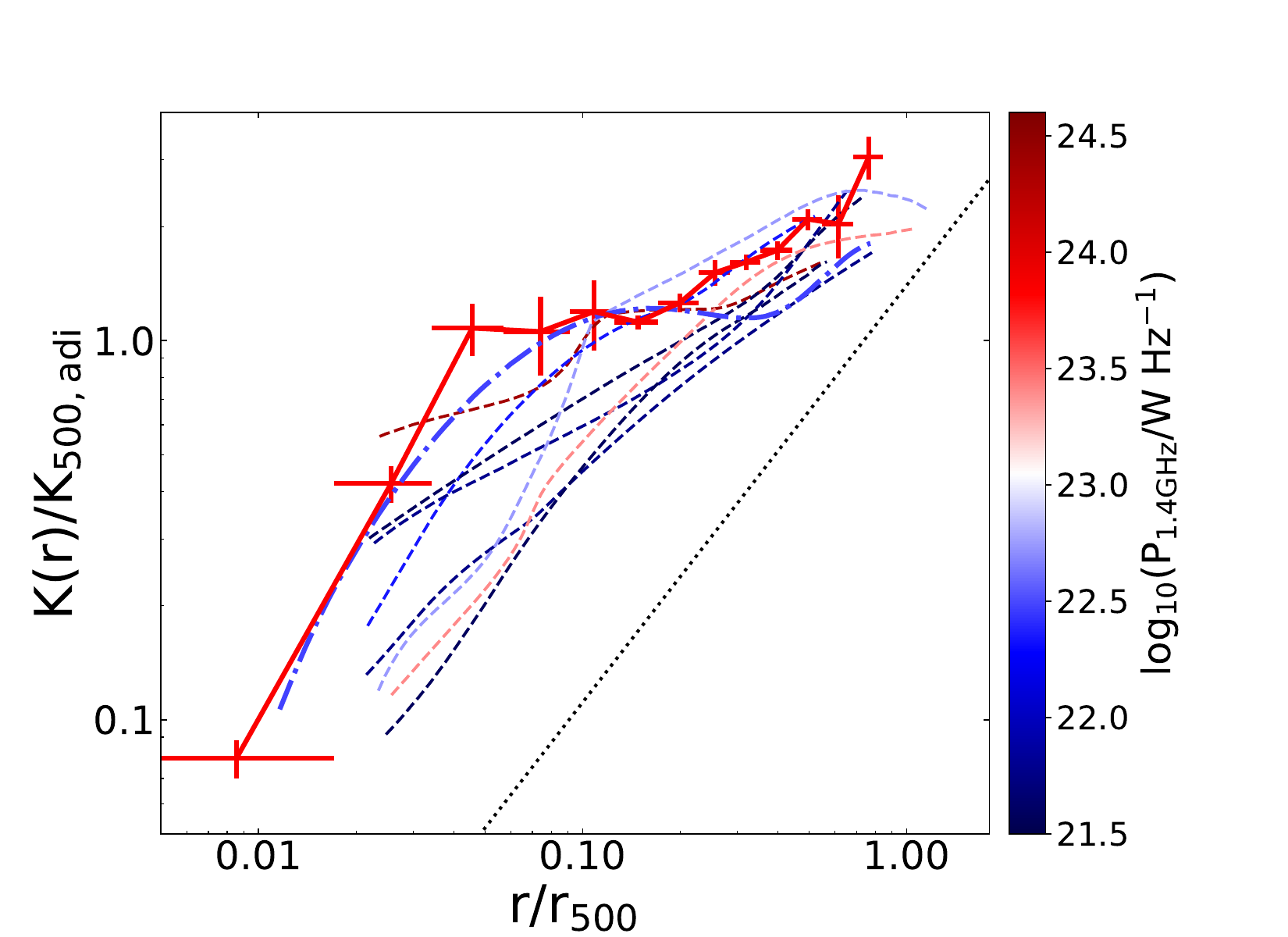} 
	\caption{Scaled entropy profile of IC~2476 (solid line) and entropy profiles of galaxy groups with similar system temperatures as IC~2476 from \citet{Sun09} (dashed lines). The colors of the lines indicate the radio power at 1.4 GHz, log$_{10}$$(P_{\rm 1.4 GHz}$), of the central BCG. Among them, galaxy group 3C~449 has the strongest radio AGN, with a radio power of $2.4\times10^{24}$ W Hz$^{-1}$ at 1.4 GHz. We also added the scaled entropy profile of galaxy group SDSSTG~4436 (dash-dotted line), which has a higher system temperature ($\sim1.85$ keV) than IC~2476 and may have experienced extreme AGN feedback as proposed by \cite{Eckert2025a}. The black dotted line represents the baseline entropy profile expected from gravitational collapse \citep{Voit05}.
\label{K_scaled}}
\end{center}
\end{figure}

\begin{figure}
\begin{center}
\hspace{-15px}
\includegraphics[scale=0.34]{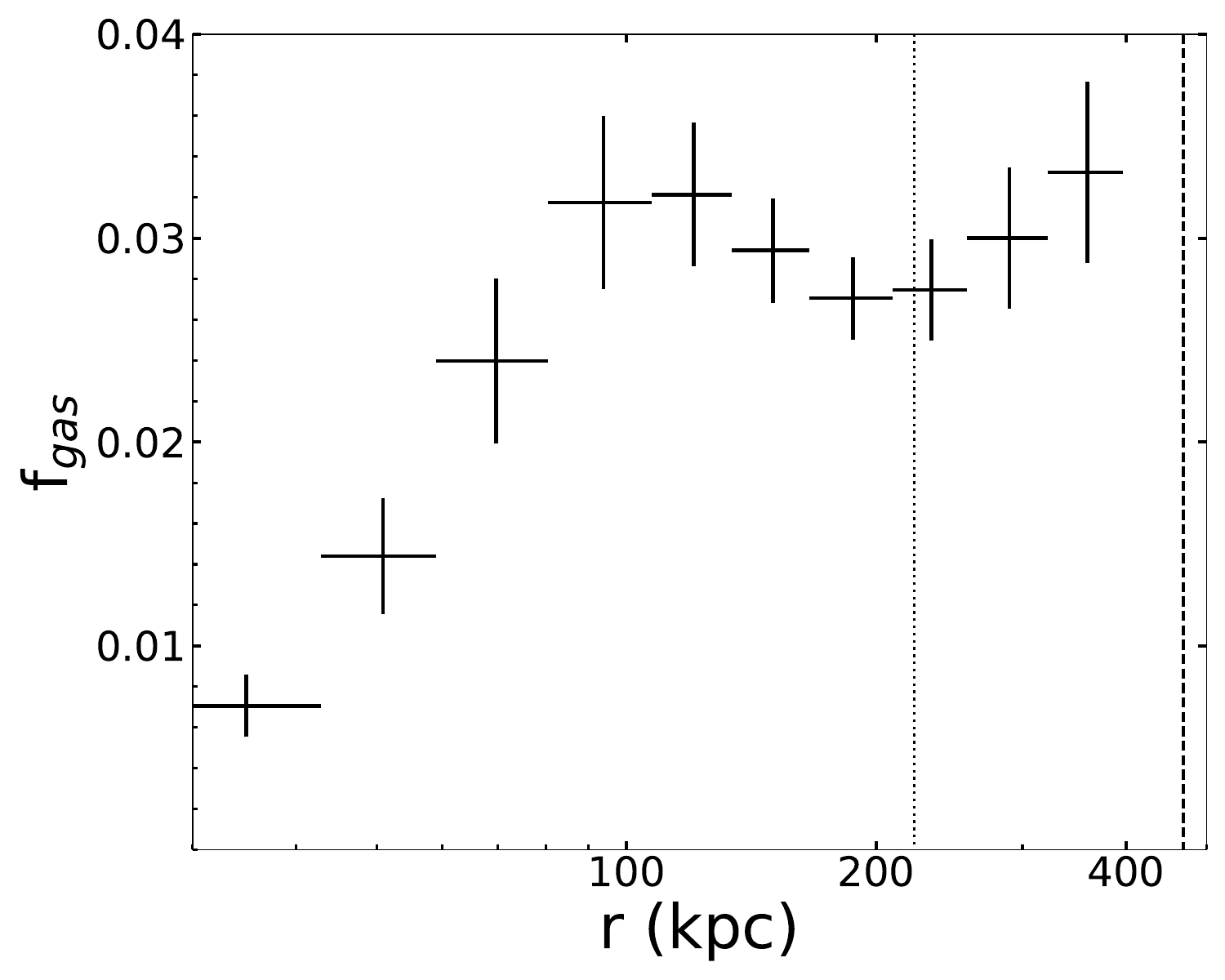} 
	\caption{Gas fraction of galaxy group IC~2476. The vertical dotted line marks $r_{2500}$, and the vertical dashed line marks $r_{500}$.
\label{fg}}
\end{center}
\end{figure}

\section{Results}
\label{Results}
\subsection{Thermal Properties}
Fig. \ref{PST} shows the radial profiles of gas pressure, entropy, and the cooling time for galaxy group IC~2476,
which are given by $P=nkT$, $S=kT/n_{e}^{2/3}$, and $t_{cool}=\frac{3nkT}{2n_{e}n_{H}\Lambda(T,Z)}$,
where $n$ is the total gas number density, $n_{e}$ and $n_{H}$ are the electron and proton number densities,
$\Lambda(T,Z)$ is the cooling function for a given temperature and metal abundance, which is calculated using the APEC model \citep{Smith2001}.
The pressure decreases monotonically with radius.
The entropy rises steeply, from $\sim10$ keV cm$^{2}$ at the center to $\sim131$ keV cm$^2$ at 20 kpc, where the profile flattens, before rising again beyond $\sim70$ kpc.
The cooling time in the innermost bin is $\sim0.8$ Gyr. 
Following the previous study \citep[e.g.,][]{Liu2024}, we defined the cooling radius as $r_{4Gyr}$, 
within which the gas cooling time is shorter than 4 Gyr, instead of
the conventionally used value of 7.7 Gyr, since the cooling radius $r_{4Gyr}$ can effectively encompass the whole of a small cool core \citep{Sun2007}. 
The cooling radius is estimated to be $\sim10$ kpc, showing that IC~2476 has a small cool core.

Fig. \ref{K_scaled} shows the scaled entropy profile, $K(r)/K_{\rm 500,adi}$, of IC~2476, where $K_{\rm 500,adi}$ is
the adiabatic entropy scale at $r_{500}$ \citep{Sun09}, defined as:
\begin{equation}
	K_{\rm 500,adi}= 342\ {\rm keV\, cm^{2}} \ (\frac{M_{500}}{10^{14} M_{\odot}})^{2/3} E(z)^{-2/3}h_{70}^{-4/3},
	\label{eq1}
\end{equation}
where $E(z)=[\Omega_{M}(1+z)^{3}+\Omega_{\Lambda}]^{1/2}$, and $h_{70}=\frac{H_{0}}{\rm 70\ km\ s^{-1} Mpc^{-1}}$.
We compared the scaled entropy of IC~2476 to other galaxy groups with similar system temperature ($kT_{500}$ in the range
0.87-1.18 keV) obtained from \citet{Sun09} in Fig. \ref{K_scaled}. IC~2476 generally has higher scaled entropy than other X-ray groups in this comparison. 
The scaled entropy of IC~2476 rises steeply up to $\sim 0.04$ $r_{500}$. It becomes flat from $0.04-0.1$ $r_{500}$, and is higher than the scaled entropy profiles
of other galaxy groups with similar system temperatures, which indicates a high injection of energy from the past 
activity of central AGN outside the central cool core. We calculated the offset between the X-ray peak and the central BCG, which is less than 1 kpc. 
We also calculated surface brightness concentration parameter, $c_{SB}$, of $\sim0.29$ using the SPA code\footnote{https://sites.google.com/site/adambmantz/code/morph14}, 
suggesting the presence of a cool core \citep{Santos2008}. The small offset between the X-ray peak and the central BCG, and the presence of cool core imply that IC~2476 
is a relaxed system, therefore, ruling out the recent merging event as the potential source of energy input.
We also compared the scaled entropy profile of IC~2476 to that of galaxy group SDSSTG~4436 with a higher system temperature ($\sim1.85$ keV), which likely experienced 
extreme AGN feedback \citep{Eckert2025a}. 
Both systems currently lack strong radio AGN activity. While IC~2476 shows a generally higher scaled entropy than SDSSTG~4436 over most radii, the scaled entropy 
profiles of two systems display similar radial trends: a steep rise in the center, a flattened profile at intermediate radii, and 
an increase toward larger radii.

We further measured the radial gas fraction profile of IC~2476, as shown in Fig. \ref{fg}. The gas mass is calculated
by integrating the gas density ($\rho=n_{e} \mu_{e} m_{p}$, where $\mu_{e}\approx1.15$ is the mean particle weight per electron, and $m_{p}$ is the
proton mass) in shells. Assuming hydrostatic equilibrium and spherical symmetry, we can calculate the gravitational acceleration 
$g$ \citep{Sarazin1988}: $g=\frac{d\Phi}{dr}=-\frac{1}{\rho}\frac{dP}{dr}$, where $\Phi$ is the gravitational potential.
We fitted the pressure profile using a generalized
NFW model \citep[e.g.,][]{Nagai2007, Arnaud2010} with $P \propto r^{-3}$ at large radii and allowed the inner slope to vary, to get a well-behaved pressure gradient. 
While massive clusters typically exhibit a steeper outer slope of $\sim-5.5$ for the pressure profile \citep{Arnaud2010}, our analysis of galaxy group IC~2476 shows that
a flatter outer slope is preferred. We got a smaller $\chi^{2}$ ($\Delta \chi^{2}\sim3.0$) when fixing the outer slope to $-3$ than to $-5.5$.
Then we estimated the enclosed total mass profile based on the gravitational acceleration.
We measured a total mass of $\sim1.75\times10^{13}$ $M_\odot$ at $r_{2500}$ and a gas fraction of $\sim0.03$, which is in agreement with the previous studies 
for galaxy groups \citep[e.g.,][]{Sun09}. 
We also verified that the choice of outer slope does not systematically bias our result. The derived mass and gas fraction remain consistent within $1\sigma$ error.

\subsection{Cool Core Remnant}
\label{coolcore}
\begin{figure}
\vspace{8px}
\begin{center}
\hspace{-10px}
\includegraphics[scale=0.32]{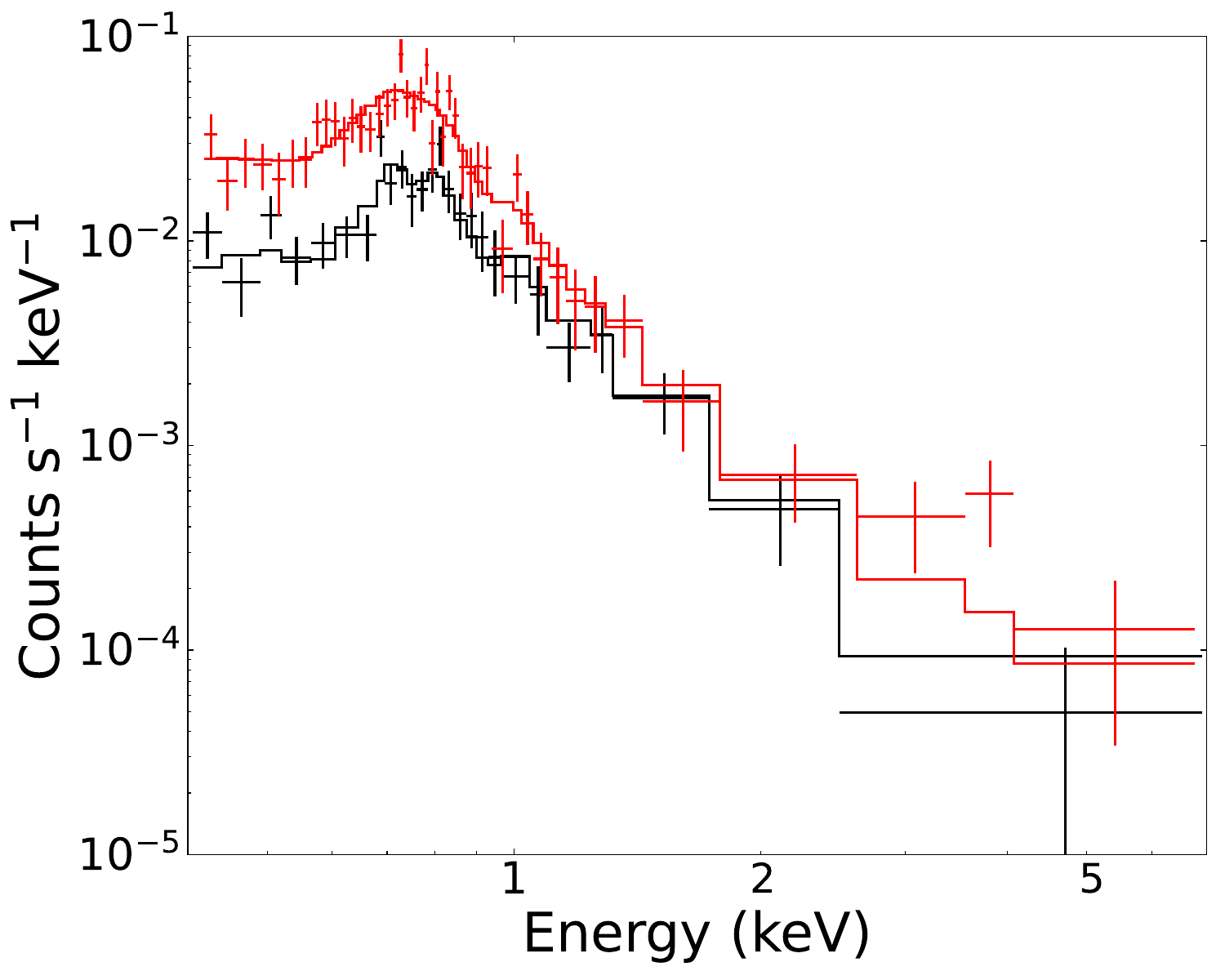} 
	\caption{Combined MOS (black) and PN (red) spectra of the central cool core. The solid lines represent the best-fitting model including a thermal component in which the abundance is allowed to vary and a power-law component (see Table~\ref{tab3} for details).
\label{Spec}}
\end{center}
\end{figure}

We then checked the properties in the central cool core.
We extracted the spectrum from the cool core region centered on the BCG within the
radius $r_{4Gyr}$. The background spectrum was extracted within the region from $r_{4Gyr}$ to 2$r_{4Gyr}$. 
The MOS/PN spectra from different observations are combined using the tool {\tt epicspeccombine}.
We fitted the combined MOS/PN spectra simultaneously. 
We first tried the model only including an absorbed thermal component. We allowed the temperature, the metallicity and the normalization to vary. 
We then tried the model including an additional power-law component for any possible residual emission from the central AGN with the power-law index fixed at 1.7. 
The metal abundance is $0.14^{+0.06}_{-0.04}$ solar. We also tried to fix it at 0.5 solar. We found that the temperature is not very sensitive to the abundance value. 
Fig. \ref{Spec} shows the combined MOS and PN spectra and the best-fitting model.
The fitting results are summarized in Table \ref{tab3}. A power-law component obviously improves the fit. 
The temperature within the cool core $r_{4Gyr}$ is
$0.45\pm{0.04}$ keV. The X-ray luminosity in the $0.5-2.0$ keV band within the cool core is $1.50\times10^{40}$ erg s$^{-1}$.
The small size, low X-ray luminosity, and high gas density of the central cool core are consistent with the remnant of a large cool core that has been disrupted by past AGN heating.

\begin{deluxetable*}{lccccccc}
	\tablecaption{Summary of Fit Results of Central Cool Core \label{tab3}}
\tablehead{
	\colhead{Model} &  \colhead{kT} & \colhead{Z}   & \colhead{$L_{\rm 0.5-2\ keV}\tablenotemark{a}$} & \colhead{Index} & \colhead{$L_{\rm 2-10\ keV}\tablenotemark{b}$}  & C-statistic/dof \\
	 \colhead{}      & \colhead{keV}     & \colhead{} & \colhead{$10^{40}$ ergs s$^{-1}$}          & \colhead{} & \colhead{$10^{39}$ ergs s$^{-1}$} & \colhead{} 
 }
\startdata
		1T &	 $0.52_{-0.03}^{+0.03}$ & $0.09_{-0.02}^{+0.02}$ & $1.73_{-0.18}^{+0.20}$ &      &                        & 97.3/87\\
	        1T + Pow &     $0.45_{-0.04}^{+0.04}$ & $0.14_{-0.04}^{+0.06}$ & $1.53_{-0.32}^{+0.30}$ & 1.7  & $3.99_{-1.32}^{+1.28}$ & 88.0/86\\
	        1T + Pow &     $0.37_{-0.02}^{+0.02}$ & $0.5$                  & $1.41_{-0.11}^{+0.11}$ & 1.7  & $5.91_{-0.95}^{+0.97}$ & 94.1/87\\
\enddata
\tablenotetext{a}{X-ray luminosity in the $0.5-2.0$ keV band of the thermal component.}
\tablenotetext{b}{X-ray luminosity in the $2.0-10.0$ keV band of the power-law component.}
\end{deluxetable*}

\begin{figure*}
\begin{center}
\vspace{-10px}
\begin{tabular}{ccc}
\includegraphics[width=2.25in]{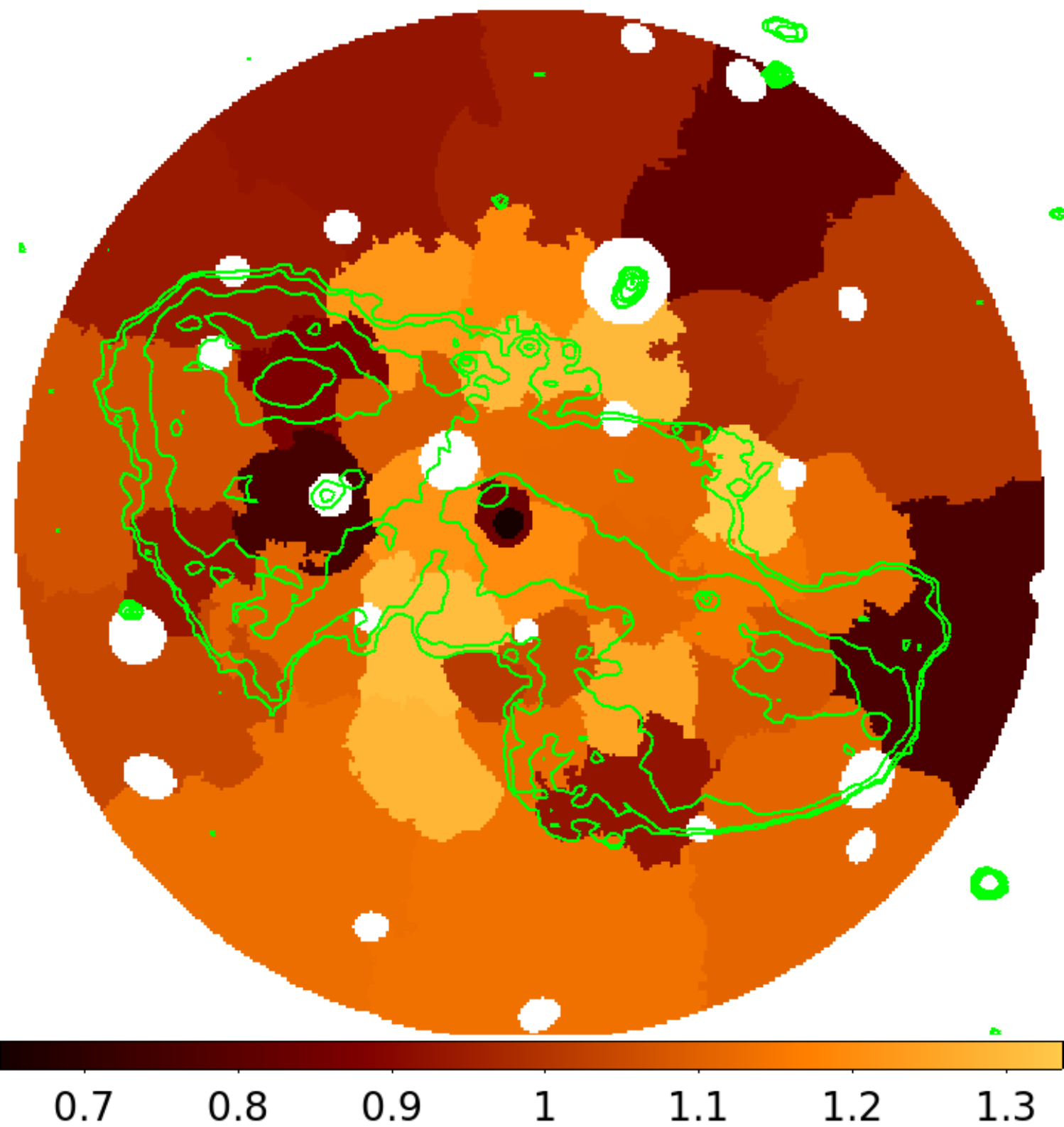} & 
\includegraphics[width=2.25in]{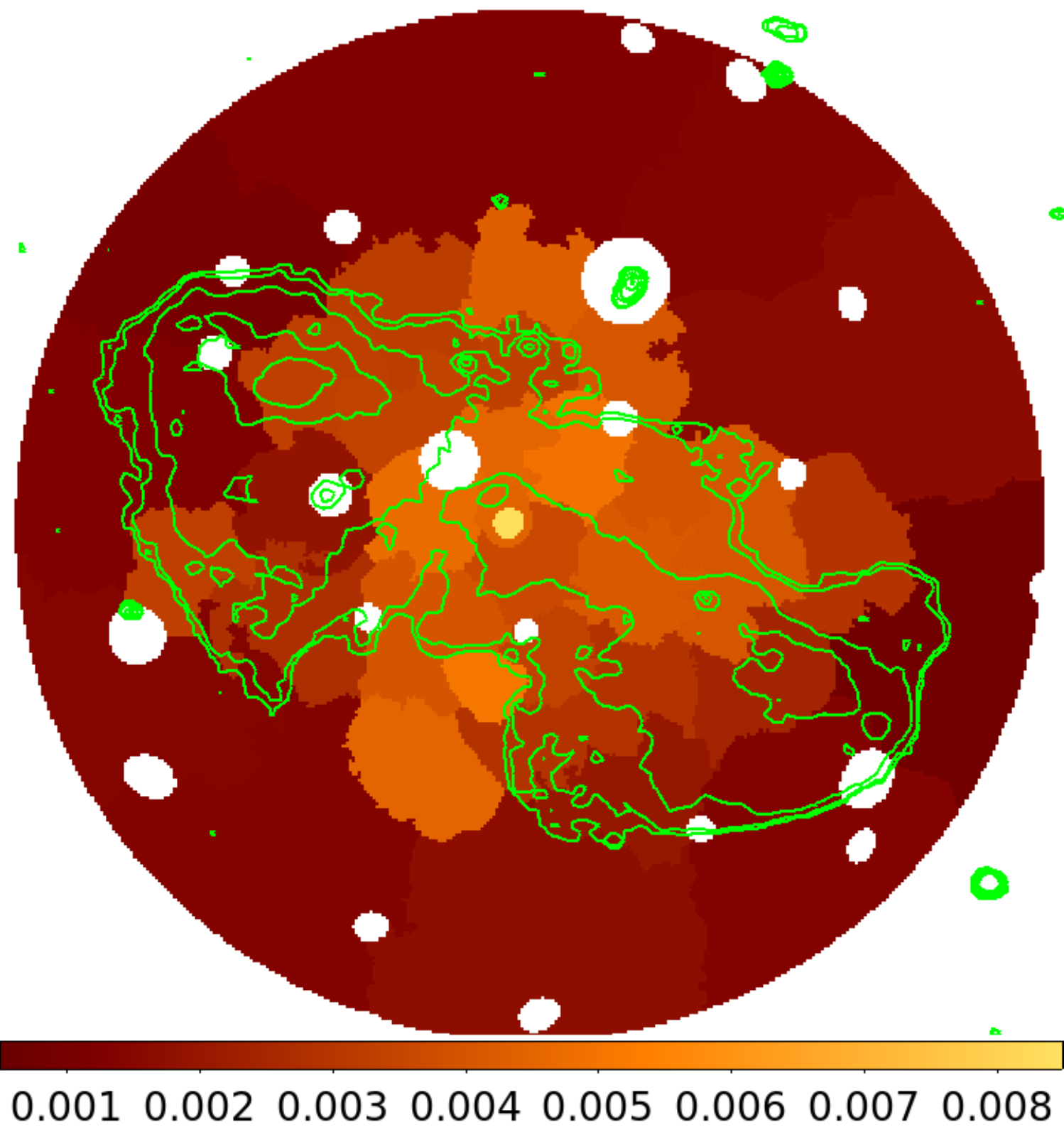}  &
\includegraphics[width=2.25in]{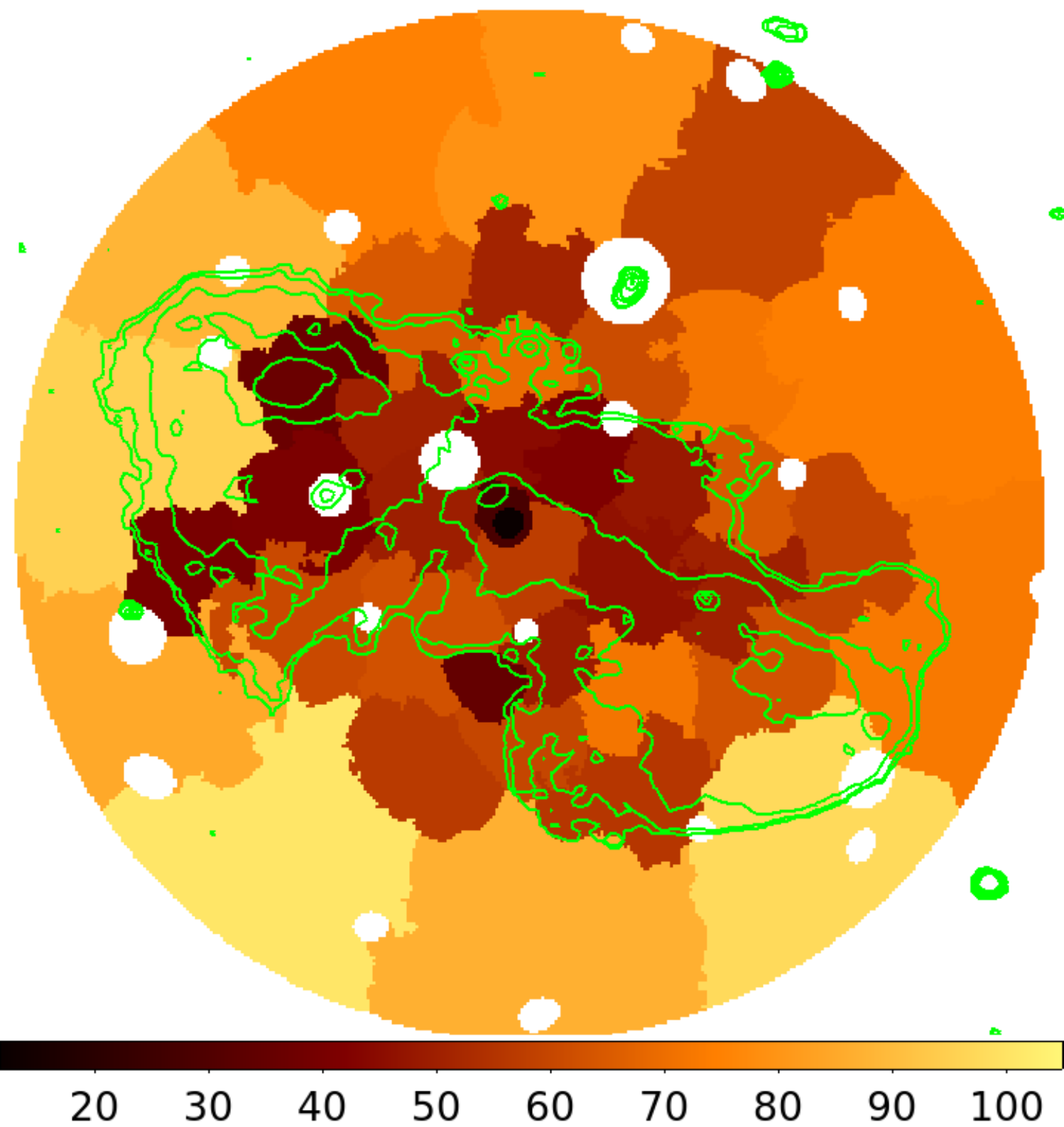} 
\end{tabular}
	\caption{Temperature map (left, in the unit of keV), pseudo pressure map in arbitrary units (middle), and pseudo entropy map in arbitrary units (right) 
	of IC~2476 within the central 7 arcmin.
		\label{map}}
\end{center}
\end{figure*}

\subsection{X-ray Cavity and AGN Heating}
In Section \ref{Analysis}, we found that there is a surface brightness depression only in the eastern wedge, coinciding
with the eastern radio lobe. However, from the X-ray image, there is no obvious boundary to the X-ray cavity, suggesting that the X-ray gas
may have mixed with the radio plasma across this boundary.
Therefore, to estimate the X-ray cavity power, we assumed an X-ray cavity within the eastern radio lobe, which is approximately
represented with a sphere at a projected distance of $\sim90$ kpc from the central BCG, and with a radius of $\sim32$ kpc, based 
on the surface brightness profile.
We estimated a total enthalpy ($H=4PV$, where $P$ is the azimuthally-averaged pressure at the radius of the cavity center) of $4.3\times10^{58}$ erg. 
The buoyant rise time of the bubble is calculated at its terminal velocity $\sim(2gV/SC)^{1/2}$, where $V$ is the volume of the bubble,
$S$ is the cross section, $C=0.75$ is the drag coefficient, and $g$ is the gravitational acceleration \citep{Churazov2001}.
We obtained a buoyant rise time of $\sim2.1\times10^{8}$ yr. Therefore, the cavity power is estimated to be
$\sim6.4\times10^{42}$ erg s$^{-1}$. Based on the $P_{radio}-P_{cavity}$ relation \citep[e.g.,][]{Cavagnolo2010}, the jet power is estimated to
be $\sim4.0\times10^{43}$ erg s$^{-1}$ and $\sim7.4\times10^{43}$ erg s$^{-1}$ from 325 MHz and 1.4 GHz radio luminosity, which are about 5.3 and 10.6 
times higher than the cavity power estimated based on the total enthalpy and
the buoyant rise time of the bubble. If we use the age of $\sim80$ Myr \citep{Turner2018} for the source, the cavity power is $\sim1.7\times10^{43}$ erg s$^{-1}$. 
Noted that, since the central engine of the radio galaxy IC~2476 has shut down and the X-ray cavity
may be dissipating, the cavity power could be significantly underestimated. The $P_{radio}-P_{cavity}$ relation may not work for remnant radio sources.

\subsection{Temperature, Pressure, and Entropy Maps}
To study the thermal structures of IC~2476, we performed a 2D spectroscopic analysis using contour binning \citep{Sanders2006}.
We generated 40 regions with a signal-to-noise ratio of 35 using the count image in the 0.4-1.3 keV band within the central 7 arcmin. 
The gas emission in each region is modeled with {\tt TBABS*APEC}, and the absorbing column density is fixed. 
With the derived temperature and normalization in each region, we then calculated the pseudo-pressure and pseudo-entropy 
(in arbitrary units) following the method
in \citet{Ubertosi2023}. The resulting temperature, pseudo-pressure, and pseudo-entropy maps are shown in Fig. \ref{map}.
These maps show that the gas in the central cool core region has a low temperature, high pressure, and the lowest entropy. We also noticed that the 
X-ray blob shown in the right panel of Fig. \ref{image} has a relatively low temperature and low entropy, suggesting that it may be uplifted or entrained from the center.

\section{Discussion}
\label{discussion}
\subsection{Low Gas Temperature in Central Cool Core}
In Section \ref{Results} we found that the gas temperature within the central cool core ($\sim10$ kpc) is below 0.50 keV,
whether or not we fixed the abundance of the thermal component used in the spectral model.
This is not usually seen in the core region of galaxy groups. \citet{Sun2009} measured the gas temperature within the central cool core
coronae of the BCGs for a large sample of galaxy groups and clusters, using the same definition of the size of cool core coronae as in this paper. In their study, the lowest
temperature measured for a central cool core coronae is 0.59 keV, which is higher than the value obtained for IC~2476.
The cooling time within the cool core corona, corresponding to the innermost data point in the right panel of Fig. \ref{PST}, is approximately $\sim0.8$ Gyr. In addition, 
IC~2476 is currently a remnant radio source with its central AGN switched off $\sim30-50$ Myr ago \citep[e.g.,][]{Shulevski2017,Turner2018}. We can estimate how much the gas 
temperature can change in the absence of heating. Assuming isobaric radiative cooling and a constant cooling time $t_{cool}=0.8$ Gyr, gas initially at $kT_{0}=0.60$ keV in 
the core of IC~2476 would cool to approximately as $T\approx T_{0}e^{-0.6t/t_{cool}}=T_{0}e^{-0.04} \sim0.58$ keV over 50 Myr, higher than the value found in the cool core of IC~2476. 
However, owing to the poor spatial resolution of \xmm\ compared to \chandra, we cannot resolve the regions closer to the very center, where the cooling time could be shorter. 
Moreover, as discussed in \citet{Liu2024}, AGN outbursts may disrupt a substantial fraction of the cool core while leaving a compact cool core corona, particularly if most of 
the jet energy is deposited outside the central $\sim10$ kpc \citep[e.g.,][]{Soker2016}. In such a scenario, the central gas could cool even before the radio source completely 
shuts down. However, the current data do not allow us to distinguish between these possibilities, and deeper multiwavelength observations will be required to assess the 
presence of multi-phase gas in the core region.

\subsection{Evolution Stage}
The central radio galaxy has bright radio lobes, however, no radio jets and a radio weak nucleus, suggesting that the radio AGN is currently in a quiescent state. 
Using radiative ageing model fitting, \citet{Shulevski2017} found that the active phase duration of IC~2476 is about 100 Myr, and the time since the shutdown is $\sim50$ Myr. 
\citet{Turner2018} used a new approach that combines spectral aging and dynamical models to estimate that the radio source
in galaxy group IC~2476 spent $50^{+14}_{-12}$ Myr in its active phase and has been quiescent for a further $28^{+6}_{-5}$ Myr.
The total jet kinetic power is $3.6\times10^{44}$ erg s$^{-1}$.
Based on the X-ray data, we derived a dynamical age (buoyant rise time) of $2.1\times10^{8}$ yr, and a power of $6.4\times10^{42}$ erg s$^{-1}$
for the X-ray cavity. The discrepancy between the radio study and our result may be due to several factors. Firstly, since the central AGN has been inactive for tens of Myr, 
it is possible that some of the radio plasma in X-ray cavity has mixed with the surrounding hot gas and dissipated. The original X-ray cavity could have been much larger 
than the one assumed in our calculation, resulting in an underestimate of the total cavity enthalpy. Secondly, the cavity power calculation is based on the buoyant rise time, 
which can be much larger than the active time for the radio jets \citep[e.g.,][]{Hardcastle2020}. In fact, the radio lobes can remain overpressured and continue expanding 
when the jets switch off \citep[e.g.,][]{Stewart2025}. Therefore, the time-averaged cavity power is smaller than the jet power estimated in \citet{Turner2018} with a much 
smaller active time of radio jet. Thirdly, \citet{Turner2018} assumed that the lobes expanded against a constant environmental pressure in the dynamical model, which is 
certainly not accurate based on the X-ray measurements. However, detailed analysis of the difference caused by the external gas pressure in their model is beyond the scope of this paper.

\citet{Sun2009} studied a large sample of galaxy groups and clusters, and in general the X-ray cool cores can be classified as
large cool cores and cool core coronae. Galaxy groups with luminous radio AGN only have small coronae. 
\citet{Liu2024} studied a sample of 108 nearby galaxy clusters and proposed a transition from large X-ray cool cores to cool core
coronae due to radio AGN heating in galaxy clusters. Using the same definition for the cool core size, we obtained a cool core
radius of $\sim10$ kpc and derived an X-ray luminosity of $1.5\times10^{40}$ erg s$^{-1}$ in the $0.5-2.0$ keV band for IC~2476. Based on its radio
power and the fact that the central engine has been shut down, we suggest that the previous powerful radio AGN outburst has disrupted
the cool core of IC~2476. Without the heating from the central AGN, the size of cool core may increase.

Recent stacking results from X-ray, kinetic Sunyaev-Zel'dovich, and weak-lensing observations suggest that AGN feedback may be even stronger than what has been adopted 
from cosmological simulations \citep[e.g.,][]{McCarthy2025,Siegel2025}. By comparing the X-ray $L_{X}-T$ relation in a sample of 44 nearby galaxy groups with the 
predictions from FLAMINGO simulations, \citet{Eckert2025b} found that highly ejective feedback model is in tension with the measurements of local galaxy groups. The X-ray 
luminosity and temperature of IC~2476
is consistent, within the scatter, with the observed $L_{X}-T$ relation of \citet{Eckert2025b}, but inconsistent with the -8$\sigma$ strong-feedback model, 
because that model would require a factor of several lower luminosity at the same temperature. This suggests that the past AGN activity of IC~2476 heated and 
redistributed the intragroup medium, efficiently disrupting the cool core while not 
ejecting too much gas to large radii.

\subsection{Pressure Balance}
Assuming equipartition between the energy contained in the magnetic field and in relativistic particles, and an electron to proton
ratio of unity, \citet{Shulevski2017}
estimated a magnetic field strength of 1.35 $\mu G$, which is equivalent to the total lobe internal pressure of $7.3\times10^{-15}$ Pa or $\sim4.5\times10^{-5}$
keV cm$^{-3}$ considering the magnetic pressure and particle pressure. Based on the X-ray measurements, the ratios of the internal lobe pressure to the external 
ICM pressures are estimated to be $\sim0.03$ at the lobe mid-point and $\sim0.07$ in the outer lobe regions.
These values suggest that the radio lobes are underpressurised relative to the surrounding hot gas, as commonly found
in low-power FR-I radio galaxies \citep[e.g.,][]{Croston2008b}.
The magnetic field estimated in \citet{Shulevski2017} is based on the method of \citet{Miley1980}, which will overestimate the magnetic field for a steep and 
curved-spectrum source. Therefore, the internal pressure could be even overestimated.
If in reality the radio lobes are close to pressure balance with the external medium, this would imply the presence of a significant non-radiating particle component, 
such as protons, with a proton to electron ratio greater than unity.

\begin{figure}
\vspace{8px}
\begin{center}
\hspace{-10px}
\includegraphics[scale=0.32]{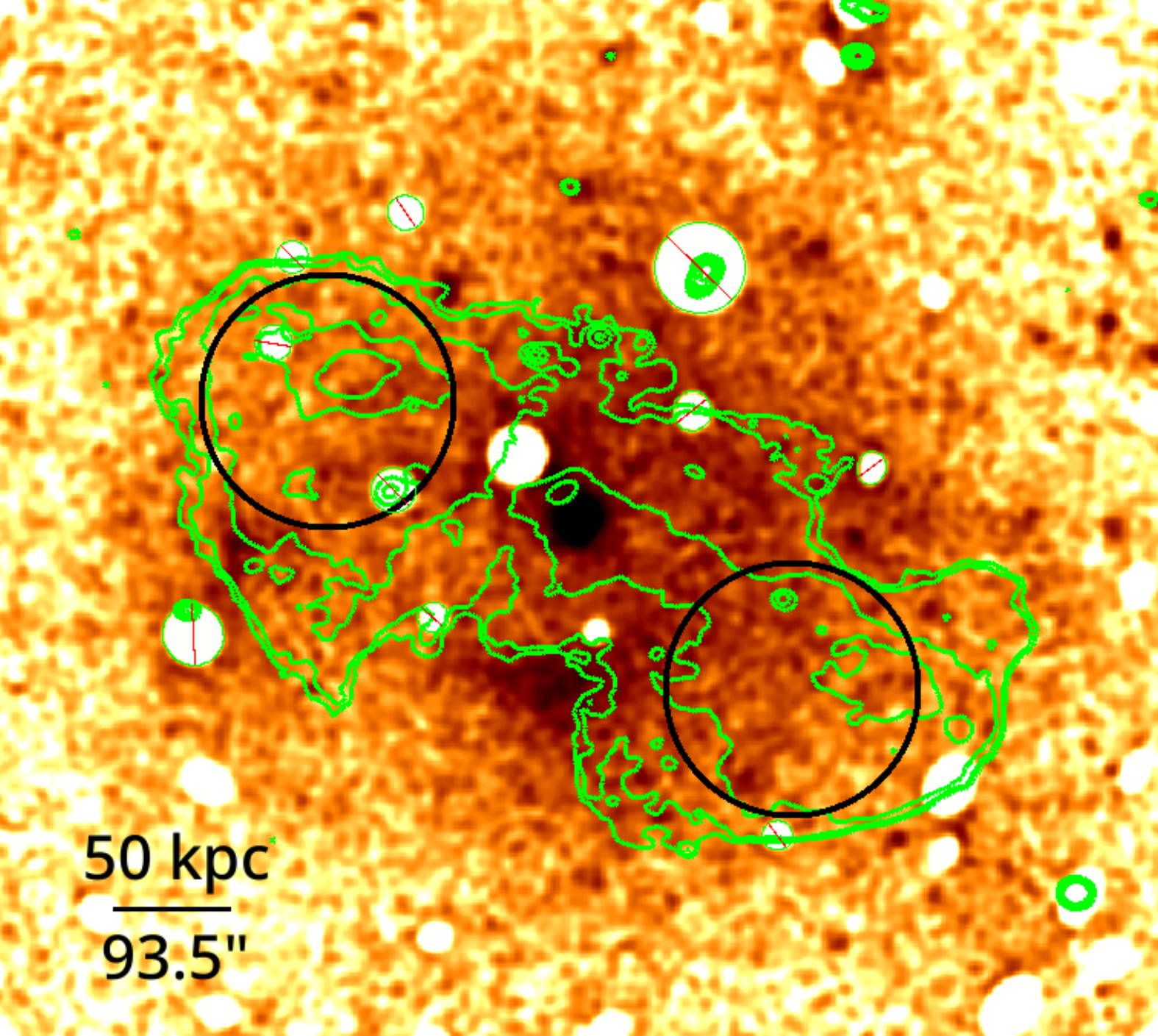} 
	\caption{Regions used to study the diffuse X-ray emission in the radio lobes of IC~2476. The radio contours are shown in green. Solid circles with a radius 
	of $53.5$ kpc are regions used for the eastern and western lobes.
\label{lobe}}
\end{center}
\end{figure}

For comparison, \citet{Ineson2017} found that in powerful FR-II radio galaxies the lobes are approximately in pressure balance at their mid-points and overpressured 
toward their outer regions relative to the external medium, and inverse Compton (IC) X-ray emission
from the lobes is commonly detected \citep[e.g.,][]{Migliori2007,Isobe2011}. 
To check for the IC X-ray emission in the lobes, we chose two circular regions with a radius of $53.5$ kpc and the same distances ($\sim116.0$ kpc) to the BCG, which 
are close to the outer parts of the eastern and western lobes, as shown in Fig. \ref{lobe}. We extracted the spectra from these two regions from each observation and 
fitted the spectra the same way as we did in each annulus for the projected temperature. We used an absorbed thermal component for the hot gas in the group and an 
additional power-law component for the possible IC X-ray emission. We found that an extra power-law component helps to improve the fit only for the eastern lobe. The 
power-law photon index is $2.50^{+0.12}_{-0.16}$, and the luminosity at 2-10 keV is $3.02\times10^{40}$ erg s$^{-1}$. The gas temperature is $0.80^{+0.07}_{-0.08}$ keV, 
lower than the projected temperature at the same distance, suggesting that it could be the up-lifted low temperature gas from the center by the radio jets. However, 
we found that a second thermal component with a higher temperature ($\sim2$ keV) can also help to fit the spectra. We cannot make a definitive conclusion as to which model 
is better. In contrast, for the western lobe, an extra power-law model does not help to improve the fit. The fitted gas temperature is $1.14\pm0.04$ keV, which is consistent
with the projected temperature.

\subsection{Low Abundance}
The exceptionally low iron abundance measured for IC 2476 is difficult to reconcile with previous studies on galaxy clusters and 
groups \citep[e.g.,][]{Rasmussen2007, Mernier2017,Gastaldello2021}, in which the iron abundance is generally higher in the central region and decreases toward 
larger radii (e.g., $\ge0.2$ solar). We carefully examined the possible
systematic uncertainties. The treatment of the sky background is critical for a low-surface-brightness system such as IC 2476. We increased the sky background normalizations by 50\% and found that the derived abundance can reach $\sim0.4$ solar (upper limit) in the outer region, whereas it remains almost unchanged in the central, source-dominated region. Another systematic uncertainty arises from the Fe-bias effect \citep{Buote00a}, in which a single-temperature fit to a multi-temperature plasma systematically underestimates the iron abundance. To address this, we fitted the spectra with two thermal components with tied abundances. We found that the best-fit value in each annulus is still below $\sim0.2$ solar, although with large error bars in some regions, e.g., the upper limit reaching $\sim0.5$ solar in the third annulus. We also tried the multi-temperature collisional-ionization equilibrium plasma emission model, {\tt gadem}, in which the distribution of emission measure is assumed to follow a Gaussian profile as a function of temperature. Due to the degeneracy between the parameters of temperature sigma ($\sigma_{T}$) and the abundance in {\tt gadem} model, we tested two temperature sigma values of $0.2$ and $0.5$. For $\sigma_{T}=0.2$, the fitted abundance is consistent with the projected values. For $\sigma_{T}=0.5$, the abundances are systematically higher in the central $\sim107$ kpc with the best-fit values of $\sim0.2-0.4$ solar, and are consistent with the projected values beyond $\sim107$ kpc. 

Recently, \citet{Hopkins2025} proposed that inverse-Compton (IC) scattering of CMB photons by low energy ($\sim0.1-1$ GeV) Cosmic Rays (CRs; lifetime of $\ge$ Gyr) can produce $\sim$ keV X-rays with nearly thermal X-ray spectra, mimicking cool core emission in groups and clusters. Therefore, a significant IC contribution would bias X-ray gas-phase metallicities low, providing a possible explanation for the exceptionally low iron abundance found for IC~2476. However, testing this scenario requires detailed modeling of the IC X-ray emission, and we leave it for further work.

In summary, if the low iron abundance measured in the intragroup medium of IC~2476 is real, it would imply that the metal enrichment history of IC~2476 differs substantially from other systems, potentially reflecting the combined effect of gas accretion, metal loss, and AGN-driven redistribution.

We also assessed systematic uncertainties by fixing the metallicity to 0.3 solar during spectral fitting. The resulting projected temperatures are highly consistent with those obtained when the metallicity is allowed to vary freely, whereas the gas densities are systematically lower. Consequently, the derived gas pressures and cooling times are systematically lower, and the gas entropies are systematically higher. However, all differences remain within the $2\sigma$ confidence level. The main conclusions of our study are unaffected.

\section{Summary and Conclusions}
\label{Conclusion}
Radio observations show that the central radio AGN of IC~2476 is currently off, although the radio lobes from the past AGN have not faded yet. 
We analyzed a deep \xmm\ observation ($\sim$180 ks) of galaxy group IC~2476 and presented
the properties of the hot gas related to the AGN feedback. The results are summarized as the following:
\begin{itemize}
	\item We find that the group has a $\sim10$ kpc cool core corona. We argue that this is all the remains of a large cool core that has been disrupted by the radio AGN feedback. 
    The temperature within the cool core corona is $0.45^{+0.04}_{-0.04}$ keV, which is lower than usually seen in the core region of galaxy groups.
    The luminosity in the 0.5-2 keV band within the cool core corona is $1.53_{-0.32}^{+0.30}\times10^{40}$ ergs cm$^{-2}$ s$^{-1}$.
    Such a low gas temperature may be related to the lack of AGN heating after the shutdown of central AGN. 
    Multi-band observations are required to see if there is multi-phase gas in the core region of this system.
    \item We measured a remarkably low iron abundance ($\le$0.2 solar) in the intragroup medium of IC~2476 from the central region out to the large radii. Such a low abundance profile may indicate inefficient enrichment and/or retention of metals in the hot intragroup medium.
	\item  The X-ray surface brightness is not symmetric in the central region. We observed a clear surface brightness depression
		in the region overlaid by the eastern radio lobe, suggesting that there may be an X-ray cavity. However, we find
		no bright rims or a clear boundary around the cavity. Since the central AGN has been inactive for $\sim50$ Myr,
		the X-ray cavity may have begun to mix with its surroundings.
        \item The scaled entropy profile of galaxy group IC~2476 rises steeply to $\sim 0.04$ $r_{500}$ and becomes flat from $0.04-0.1$ $r_{500}$. The entropy plateau of IC~2476 is higher than the scaled entropy profiles of other galaxy groups with similar system temperatures, which indicates a high injection of energy from the past activity of central AGN outside the central cool core.
	\item  The internal pressure of the radio lobe is less than the external pressure of the hot ICM, consistent with what is found in low-power radio galaxies and implying the presence of protons. This is consistent with the fact that IC~2476 is a remnant FR-II radio galaxy, whose jets have now switched off, allowing the remnant lobes to mix with and entrain material from the surrounding intragroup medium.
\end{itemize}

Our study shows that the galaxy group IC~2476 provides an important case showing that AGN feedback is capable of destroying a group cool core, highlighting the long-lasting impact of AGN activity in galaxy groups despite the moderate present-day radio luminosity. 
The disturbed thermal structure, compact cool core corona, and the elevated central entropy suggest that IC~2476 is currently observed in a late post-outburst or quiescent evolutionary stage, in which the imprint of a previous, more powerful AGN episode remains visible long after the central engine has shut down. 
In the absence of renewed energetic feedback, radiative cooling of the intragroup medium is expected to proceed gradually, potentially allowing a large cool core to re-establish. IC~2476, therefore, illustrates that even relatively modest AGN outburst can have lasting consequences in low-mass systems, where the intragroup medium is more susceptible
to heating, and the importance of feedback in shaping the thermal evolution of galaxy groups.
These results are in general agreement with the recent results on the fossil group SDSSTG~4436 \citep{Eckert2025a}, where giant AGN outbursts have been shown to be energetic enough to unbind gas particles and disrupt cool cores, further reinforcing the critical role of feedback in low‑mass systems \citep[also see past reviews by e.g.,][]{Sun12,Eckert2021}.

\begin{acknowledgments}
	We thank the referee for important comments and suggestions that improved this manuscript.
    This work is supported by the National Key Research and Development Program of China (2023YFA1608100), the National Science Foundation of China (12233005 and 12173088), the China Manned Space Program with grants Nos. CMS-CSST-2025-A08 and CMS-CSST-2025-A20. MB acknowledges financial support from Next Generation EU funds within the National Recovery and Resilience Plan (PNRR), Mission 4 – Education and Research, Component 2 – From Research to Business (M4C2), Investment Line 3.1 – Strengthening and creation of Research Infrastructures, Project IR0000034 – ``STILES – Strengthening the Italian Leadership in ELT and SKA”, from INAF under the Large GO 2024 funding scheme (project ``MeerKAT and Euclid Team up: Exploring the galaxy-halo connection at cosmic noon”) and the Mini Grant 2023 funding scheme (project ‘Low radio frequencies as a probe of AGN jet feedback at low and high redshift’). XZ acknowledges support by Office of Science and Technology, Shanghai Municipal Government (grant Nos. 24DX1400100, ZJ2023-ZD-001).
LOFAR, the Low Frequency Array designed and constructed by ASTRON,
has facilities in several countries, that are owned by various
parties (each with their own funding sources), and that are
collectively operated by the International LOFAR Telescope
(ILT) foundation under a joint scientific policy.
\end{acknowledgments}

\begin{figure}
\begin{center}
\hspace{-20px}
\includegraphics[scale=0.26]{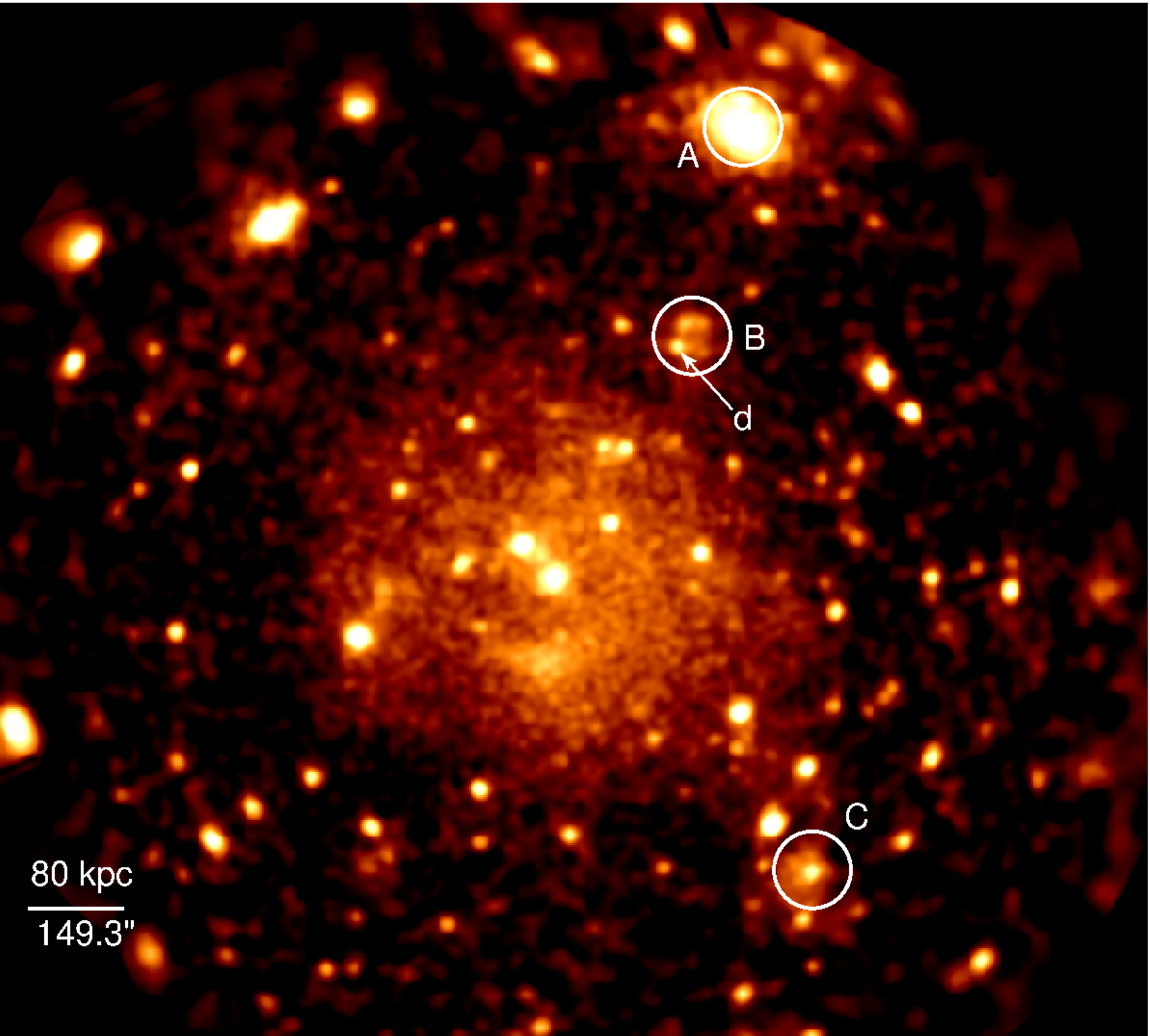} 
\caption{Combined background-subtracted, exposure-corrected \xmm\ image with three background X-ray extended emission marked as
         A, B, and C. Source A is galaxy cluster WHL J092730.0+301045 with a redshift of 0.303, whose X-ray flux in 0.5-2.0 keV
	 band is $\sim2\times10^{-13}$ erg s$^{-1}$ cm$^{-2}$ (SWIFT). Source B possibly corresponds to galaxy cluster WHL J092735.7+300511 at the redshift of 0.2098. The point source d with bright X-ray emission is a background quasar at the redshift of 2.255. Source C possibly corresponds to galaxy cluster WHL J092723.3+294959 at the redshift of 0.302.
\label{fig_a1}}
\end{center}
\end{figure}

\appendix
\label{appendix}
With the deep \xmm\ observation, we also found some background galaxy groups/clusters with extended X-ray emission within
the FOV. In Fig. \ref{fig_a1} we show the combined background-subtract, exposure-corrected flat \xmm\ image, and list the
background clusters. Source A is galaxy cluster WHL J092730.0+301045 with the redshift of 0.303. The X-ray flux in 0.5-2.0 keV
band is $\sim2\times10^{-13}$ erg s$^{-1}$ cm$^{-2}$ from SWIFT observation \citep{LiuT2015}.
Source B possibly corresponds to galaxy cluster WHL J092735.7+300511 at the redshift of 0.2098. The point source d with bright X-ray emission is a background quasar at the redshift of 2.255. Source C possibly corresponds to galaxy cluster WHL J092723.3+294959 at the redshift of 0.302.

\bibliography{my}{}

\begin{thebibliography}{}
\expandafter\ifx\csname natexlab\endcsname\relax\def\natexlab#1{#1}\fi
\providecommand{\url}[1]{\href{#1}{#1}}
\providecommand{\dodoi}[1]{doi:~\href{http://doi.org/#1}{\nolinkurl{#1}}}
\providecommand{\doeprint}[1]{\href{http://ascl.net/#1}{\nolinkurl{http://ascl.net/#1}}}
\providecommand{\doarXiv}[1]{\href{https://arxiv.org/abs/#1}{\nolinkurl{https://arxiv.org/abs/#1}}}

\bibitem[{K.~A. {Arnaud}(1996){Arnaud}}]{xspec}
{Arnaud}, K.~A. 1996, \bibinfo{title}{{XSPEC: The First Ten Years},} in ASP
  Conf. Ser. 101: Astronomical Data Analysis Software and Systems V, Vol.~5,
  17.
\newblock
  \url{http://adsabs.harvard.edu/cgi-bin/nph-bib_query?bibcode=1996adass...5...17A&db_key=AST}

\bibitem[{M. {Arnaud} {et~al.}(2010){Arnaud}, {Pratt}, {Piffaretti},
  {B{\"o}hringer}, {Croston}, \& {Pointecouteau}}]{Arnaud2010}
{Arnaud}, M., {Pratt}, G.~W., {Piffaretti}, R., {et~al.} 2010,
  \bibinfo{title}{{The universal galaxy cluster pressure profile from a
  representative sample of nearby systems (REXCESS) and the Y$_{SZ}$ -
  M$_{500}$ relation},} \aap, 517, A92, \dodoi{10.1051/0004-6361/200913416}

\bibitem[{M. {Asplund} {et~al.}(2009){Asplund}, {Grevesse}, {Sauval}, \&
  {Scott}}]{aspl}
{Asplund}, M., {Grevesse}, N., {Sauval}, A.~J., \& {Scott}, P. 2009,
  \bibinfo{title}{{The Chemical Composition of the Sun},} \araa, 47, 481,
  \dodoi{10.1146/annurev.astro.46.060407.145222}

\bibitem[{M.~C. {Begelman} \& B.~B. {Nath}(2005){Begelman} \&
  {Nath}}]{Begelman05}
{Begelman}, M.~C., \& {Nath}, B.~B. 2005, \bibinfo{title}{{Self-regulated black
  hole accretion, the M-{$\sigma$} relation and the growth of bulges in
  galaxies},} \mnras, 361, 1387, \dodoi{10.1111/j.1365-2966.2005.09249.x}

\bibitem[{D.~A. {Buote}(2000){Buote}}]{Buote00a}
{Buote}, D.~A. 2000, \bibinfo{title}{{X-ray evidence for multiphase hot gas
  with nearly solar Fe abundances in the brightest groups of galaxies},}
  \mnras, 311, 176, \dodoi{10.1046/j.1365-8711.2000.03046.x}

\bibitem[{A. {Capetti} {et~al.}(2002){Capetti}, {Celotti}, {Chiaberge}, {de
  Ruiter}, {Fanti}, {Morganti}, \& {Parma}}]{Capetti2002}
{Capetti}, A., {Celotti}, A., {Chiaberge}, M., {et~al.} 2002,
  \bibinfo{title}{{The HST survey of the B2 sample of radio-galaxies: Optical
  nuclei and the FR I/BL Lac unified scheme},} \aap, 383, 104,
  \dodoi{10.1051/0004-6361:20011714}

\bibitem[{K.~W. {Cavagnolo} {et~al.}(2009){Cavagnolo}, {Donahue}, {Voit}, \&
  {Sun}}]{Cavagnolo09}
{Cavagnolo}, K.~W., {Donahue}, M., {Voit}, G.~M., \& {Sun}, M. 2009,
  \bibinfo{title}{{Intracluster Medium Entropy Profiles for a Chandra Archival
  Sample of Galaxy Clusters},} \apjs, 182, 12,
  \dodoi{10.1088/0067-0049/182/1/12}

\bibitem[{K.~W. {Cavagnolo} {et~al.}(2010){Cavagnolo}, {McNamara}, {Nulsen},
  {Carilli}, {Jones}, \& {B{\^i}rzan}}]{Cavagnolo2010}
{Cavagnolo}, K.~W., {McNamara}, B.~R., {Nulsen}, P.~E.~J., {et~al.} 2010,
  \bibinfo{title}{{A Relationship Between AGN Jet Power and Radio Power},}
  \apj, 720, 1066, \dodoi{10.1088/0004-637X/720/2/1066}

\bibitem[{E. {Churazov} {et~al.}(2001){Churazov}, {Br{\"u}ggen}, {Kaiser},
  {B{\"o}hringer}, \& {Forman}}]{Churazov2001}
{Churazov}, E., {Br{\"u}ggen}, M., {Kaiser}, C.~R., {B{\"o}hringer}, H., \&
  {Forman}, W. 2001, \bibinfo{title}{{Evolution of Buoyant Bubbles in M87},}
  \apj, 554, 261, \dodoi{10.1086/321357}

\bibitem[{R.~A. {Cordey}(1987){Cordey}}]{Cordey1987}
{Cordey}, R.~A. 1987, \bibinfo{title}{{IC 2476 : a possible relic radio
  galaxy.},} \mnras, 227, 695, \dodoi{10.1093/mnras/227.3.695}

\bibitem[{J.~H. {Croston} {et~al.}(2008){Croston}, {Hardcastle}, {Birkinshaw},
  {Worrall}, \& {Laing}}]{Croston2008b}
{Croston}, J.~H., {Hardcastle}, M.~J., {Birkinshaw}, M., {Worrall}, D.~M., \&
  {Laing}, R.~A. 2008, \bibinfo{title}{{An XMM-Newton study of the
  environments, particle content and impact of low-power radio galaxies},}
  \mnras, 386, 1709, \dodoi{10.1111/j.1365-2966.2008.13162.x}

\bibitem[{D.~J. {Croton} {et~al.}(2006){Croton}, {Springel}, {White}, {De
  Lucia}, {Frenk}, {Gao}, {Jenkins}, {Kauffmann}, {Navarro}, \&
  {Yoshida}}]{Croton06}
{Croton}, D.~J., {Springel}, V., {White}, S.~D.~M., {et~al.} 2006,
  \bibinfo{title}{{The many lives of active galactic nuclei: cooling flows,
  black holes and the luminosities and colours of galaxies},} \mnras, 365, 11,
  \dodoi{10.1111/j.1365-2966.2005.09675.x}

\bibitem[{S. {Dasadia} {et~al.}(2016){Dasadia}, {Sun}, {Morandi}, {Sarazin},
  {Clarke}, {Nulsen}, {Massaro}, {Roediger}, {Harris}, \& {Forman}}]{Dasadia16}
{Dasadia}, S., {Sun}, M., {Morandi}, A., {et~al.} 2016,
  \bibinfo{title}{{Shocking features in the merging galaxy cluster
  RXJ0334.2-0111},} \mnras, 458, 681, \dodoi{10.1093/mnras/stw291}

\bibitem[{M. {Donahue} \& G.~M. {Voit}(2022){Donahue} \& {Voit}}]{Donahue2022}
{Donahue}, M., \& {Voit}, G.~M. 2022, \bibinfo{title}{{Baryon cycles in the
  biggest galaxies},} \physrep, 973, 1, \dodoi{10.1016/j.physrep.2022.04.005}

\bibitem[{D. {Eckert} {et~al.}(2020){Eckert}, {Finoguenov}, {Ghirardini},
  {Grandis}, {Kaefer}, {Sanders}, \& {Ramos-Ceja}}]{Eckert2020}
{Eckert}, D., {Finoguenov}, A., {Ghirardini}, V., {et~al.} 2020,
  \bibinfo{title}{{Low-scatter galaxy cluster mass proxies for the eROSITA
  all-sky survey},} The Open Journal of Astrophysics, 3, 12,
  \dodoi{10.21105/astro.2009.13944}

\bibitem[{D. {Eckert} {et~al.}(2021){Eckert}, {Gaspari}, {Gastaldello}, {Le
  Brun}, \& {O'Sullivan}}]{Eckert2021}
{Eckert}, D., {Gaspari}, M., {Gastaldello}, F., {Le Brun}, A. M.~C., \&
  {O'Sullivan}, E. 2021, \bibinfo{title}{{Feedback from Active Galactic Nuclei
  in Galaxy Groups},} Universe, 7, 142, \dodoi{10.3390/universe7050142}

\bibitem[{D. {Eckert} {et~al.}(2025{\natexlab{a}}){Eckert}, {Gastaldello},
  {Lovisari}, {McGee}, {Pasini}, {Brienza}, {Kolokythas}, {O'Sullivan},
  {Simionescu}, {Sun}, {Ayromlou}, {Bourne}, {Chen}, {Cui}, {Ettori},
  {Finoguenov}, {Gozaliasl}, {Kale}, {Mernier}, {Oppenheimer},
  {Schellenberger}, {Seppi}, \& {Tempel}}]{Eckert2025a}
{Eckert}, D., {Gastaldello}, F., {Lovisari}, L., {et~al.} 2025{\natexlab{a}},
  \bibinfo{title}{{Extreme AGN feedback in the fossil galaxy group SDSSTG
  4436},} \aap, 701, A127, \dodoi{10.1051/0004-6361/202555212}

\bibitem[{D. {Eckert} {et~al.}(2025{\natexlab{b}}){Eckert}, {Seppi},
  {Braspenning}, {Finoguenov}, {Gastaldello}, {Lovisari}, {O'Sullivan},
  {Ettori}, {Oppenheimer}, {Bourne}, {Kim}, {Sun}, {Khalil}, {Gozaliasl},
  {Bahar}, {Ghirardini}, {Cui}, {Kolokythas}, \& {McGee}}]{Eckert2025b}
{Eckert}, D., {Seppi}, R., {Braspenning}, J., {et~al.} 2025{\natexlab{b}},
  \bibinfo{title}{{The impact of strong feedback on galaxy group scaling
  relations},} arXiv e-prints, arXiv:2512.04203,
  \dodoi{10.48550/arXiv.2512.04203}

\bibitem[{A.~C. {Fabian}(2012){Fabian}}]{Fabian12}
{Fabian}, A.~C. 2012, \bibinfo{title}{{Observational Evidence of Active
  Galactic Nuclei Feedback},} \araa, 50, 455,
  \dodoi{10.1146/annurev-astro-081811-125521}

\bibitem[{M. {Gaspari} {et~al.}(2020){Gaspari}, {Tombesi}, \&
  {Cappi}}]{Gaspari2020}
{Gaspari}, M., {Tombesi}, F., \& {Cappi}, M. 2020, \bibinfo{title}{{Linking
  macro-, meso- and microscales in multiphase AGN feeding and feedback},}
  Nature Astronomy, 4, 10, \dodoi{10.1038/s41550-019-0970-1}

\bibitem[{F. {Gastaldello} {et~al.}(2021){Gastaldello}, {Simionescu},
  {Mernier}, {Biffi}, {Gaspari}, {Sato}, \& {Matsushita}}]{Gastaldello2021}
{Gastaldello}, F., {Simionescu}, A., {Mernier}, F., {et~al.} 2021,
  \bibinfo{title}{{The Metal Content of the Hot Atmospheres of Galaxy Groups},}
  Universe, 7, 208, \dodoi{10.3390/universe7070208}

\bibitem[{S. {Giodini} {et~al.}(2010){Giodini}, {Smol{\v{c}}i{\'c}},
  {Finoguenov}, {Boehringer}, {B{\^\i}rzan}, {Zamorani}, {Oklop{\v{c}}i{\'c}},
  {Pierini}, {Pratt}, {Schinnerer}, {Massey}, {Koekemoer}, {Salvato},
  {Sanders}, {Kartaltepe}, \& {Thompson}}]{Giodini2010}
{Giodini}, S., {Smol{\v{c}}i{\'c}}, V., {Finoguenov}, A., {et~al.} 2010,
  \bibinfo{title}{{Radio Galaxy Feedback in X-ray-selected Groups from COSMOS:
  The Effect on the Intracluster Medium},} \apj, 714, 218,
  \dodoi{10.1088/0004-637X/714/1/218}

\bibitem[{M.~J. {Hardcastle} \& J.~H. {Croston}(2020){Hardcastle} \&
  {Croston}}]{Hardcastle2020}
{Hardcastle}, M.~J., \& {Croston}, J.~H. 2020, \bibinfo{title}{{Radio galaxies
  and feedback from AGN jets},} \nar, 88, 101539,
  \dodoi{10.1016/j.newar.2020.101539}

\bibitem[{C.~M. {Harrison} \& C. {Ramos Almeida}(2024){Harrison} \& {Ramos
  Almeida}}]{Harrison2024}
{Harrison}, C.~M., \& {Ramos Almeida}, C. 2024, \bibinfo{title}{{Observational
  Tests of Active Galactic Nuclei Feedback: An Overview of Approaches and
  Interpretation},} Galaxies, 12, 17, \dodoi{10.3390/galaxies12020017}

\bibitem[{T.~M. {Heckman} \& P.~N. {Best}(2023){Heckman} \&
  {Best}}]{Heckman2023}
{Heckman}, T.~M., \& {Best}, P.~N. 2023, \bibinfo{title}{{A Global Inventory of
  Feedback},} Galaxies, 11, 21, \dodoi{10.3390/galaxies11010021}

\bibitem[{J. {Hlavacek-Larrondo} {et~al.}(2022){Hlavacek-Larrondo}, {Li}, \&
  {Churazov}}]{Hlavacek2022}
{Hlavacek-Larrondo}, J., {Li}, Y., \& {Churazov}, E. 2022, \bibinfo{title}{{AGN
  Feedback in Groups and Clusters of Galaxies},} arXiv e-prints,
  arXiv:2206.00098, \dodoi{10.48550/arXiv.2206.00098}

\bibitem[{P.~F. {Hopkins} {et~al.}(2025){Hopkins}, {Quataert}, {Silich},
  {Sayers}, {Ponnada}, \& {Sands}}]{Hopkins2025}
{Hopkins}, P.~F., {Quataert}, E., {Silich}, E.~M., {et~al.} 2025,
  \bibinfo{title}{{Cosmic Rays Masquerading as Cool Cores: An Inverse-Compton
  Origin for Cool Core Cluster Emission},} The Open Journal of Astrophysics, 8,
  54053, \dodoi{10.33232/001c.154053}

\bibitem[{J. {Ineson} {et~al.}(2017){Ineson}, {Croston}, {Hardcastle}, \&
  {Mingo}}]{Ineson2017}
{Ineson}, J., {Croston}, J.~H., {Hardcastle}, M.~J., \& {Mingo}, B. 2017,
  \bibinfo{title}{{A representative survey of the dynamics and energetics of FR
  II radio galaxies},} \mnras, 467, 1586, \dodoi{10.1093/mnras/stx189}

\bibitem[{N. {Isobe} {et~al.}(2011){Isobe}, {Seta}, {Gandhi}, \&
  {Tashiro}}]{Isobe2011}
{Isobe}, N., {Seta}, H., {Gandhi}, P., \& {Tashiro}, M.~S. 2011,
  \bibinfo{title}{{Suzaku Diagnostics of the Energetics in the Lobes of the
  Giant Radio Galaxy 3C 35},} \apj, 727, 82, \dodoi{10.1088/0004-637X/727/2/82}

\bibitem[{M. {Jamrozy} {et~al.}(2004){Jamrozy}, {Klein}, {Mack}, {Gregorini},
  \& {Parma}}]{Jamrozy2004}
{Jamrozy}, M., {Klein}, U., {Mack}, K.~H., {Gregorini}, L., \& {Parma}, P.
  2004, \bibinfo{title}{{Spectral ageing in the relic radio galaxy B2
  0924+30},} \aap, 427, 79, \dodoi{10.1051/0004-6361:20048056}

\bibitem[{G.~A. {Kriss} {et~al.}(1983){Kriss}, {Cioffi}, \&
  {Canizares}}]{Kriss1983}
{Kriss}, G.~A., {Cioffi}, D.~F., \& {Canizares}, C.~R. 1983,
  \bibinfo{title}{{The X-ray emitting gas in poor clusters with central
  dominant galaxies.},} \apj, 272, 439, \dodoi{10.1086/161311}

\bibitem[{K.~D. {Kuntz} \& S.~L. {Snowden}(2008){Kuntz} \&
  {Snowden}}]{Kuntz2008}
{Kuntz}, K.~D., \& {Snowden}, S.~L. 2008, \bibinfo{title}{{The EPIC-MOS
  particle-induced background spectra},} \aap, 478, 575,
  \dodoi{10.1051/0004-6361:20077912}

\bibitem[{T. {Liu} {et~al.}(2015){Liu}, {Tozzi}, {Tundo}, {Moretti}, {Rosati},
  {Wang}, {Tagliaferri}, {Campana}, \& {Giavalisco}}]{LiuT2015}
{Liu}, T., {Tozzi}, P., {Tundo}, E., {et~al.} 2015, \bibinfo{title}{{The Swift
  X-Ray Telescope Cluster Survey. III. Cluster Catalog from 2005-2012 Archival
  Data},} \apjs, 216, 28, \dodoi{10.1088/0067-0049/216/2/28}

\bibitem[{W. {Liu} {et~al.}(2017){Liu}, {Chiao}, {Collier}, {Cravens},
  {Galeazzi}, {Koutroumpa}, {Kuntz}, {Lallement}, {Lepri}, {McCammon},
  {Morgan}, {Porter}, {Snowden}, {Thomas}, {Uprety}, {Ursino}, \&
  {Walsh}}]{Liu2017}
{Liu}, W., {Chiao}, M., {Collier}, M.~R., {et~al.} 2017, \bibinfo{title}{{The
  Structure of the Local Hot Bubble},} \apj, 834, 33,
  \dodoi{10.3847/1538-4357/834/1/33}

\bibitem[{W. {Liu} {et~al.}(2019){Liu}, {Sun}, {Nulsen}, {Clarke}, {Sarazin},
  {Forman}, {Gaspari}, {Giacintucci}, {Lal}, \& {Edge}}]{Liu_3C88}
{Liu}, W., {Sun}, M., {Nulsen}, P., {et~al.} 2019, \bibinfo{title}{{AGN
  feedback in galaxy group 3C 88: cavities, shock, and jet reorientation},}
  \mnras, 484, 3376, \dodoi{10.1093/mnras/stz229}

\bibitem[{W. {Liu} {et~al.}(2024){Liu}, {Sun}, {Voit}, {Lal}, {Nulsen},
  {Gaspari}, {Sarazin}, {Ehlert}, \& {Zheng}}]{Liu2024}
{Liu}, W., {Sun}, M., {Voit}, G.~M., {et~al.} 2024, \bibinfo{title}{{X-ray cool
  core remnants heated by strong radio AGN feedback},} \mnras, 531, 2063,
  \dodoi{10.1093/mnras/stae1285}

\bibitem[{L. {Lovisari} {et~al.}(2015){Lovisari}, {Reiprich}, \&
  {Schellenberger}}]{Lovisari2015}
{Lovisari}, L., {Reiprich}, T.~H., \& {Schellenberger}, G. 2015,
  \bibinfo{title}{{Scaling properties of a complete X-ray selected galaxy group
  sample},} \aap, 573, A118, \dodoi{10.1051/0004-6361/201423954}

\bibitem[{I.~G. {McCarthy} {et~al.}(2025){McCarthy}, {Amon}, {Schaye},
  {Schaan}, {Angulo}, {Salcido}, {Schaller}, {Bigwood}, {Elbers}, {Kugel},
  {Helly}, {Forouhar Moreno}, {Frenk}, {McGibbon}, {Ondaro-Mallea}, \& {van
  Daalen}}]{McCarthy2025}
{McCarthy}, I.~G., {Amon}, A., {Schaye}, J., {et~al.} 2025,
  \bibinfo{title}{{FLAMINGO: combining kinetic SZ effect and
  galaxy{\textendash}galaxy lensing measurements to gauge the impact of
  feedback on large-scale structure},} \mnras, 540, 143,
  \dodoi{10.1093/mnras/staf731}

\bibitem[{B.~R. {McNamara} \& P.~E.~J. {Nulsen}(2007){McNamara} \&
  {Nulsen}}]{McNamara07}
{McNamara}, B.~R., \& {Nulsen}, P.~E.~J. 2007, \bibinfo{title}{{Heating Hot
  Atmospheres with Active Galactic Nuclei},} \araa, 45, 117,
  \dodoi{10.1146/annurev.astro.45.051806.110625}

\bibitem[{F. {Mernier} {et~al.}(2017){Mernier}, {de Plaa}, {Kaastra}, {Zhang},
  {Akamatsu}, {Gu}, {Kosec}, {Mao}, {Pinto}, {Reiprich}, {Sanders},
  {Simionescu}, \& {Werner}}]{Mernier2017}
{Mernier}, F., {de Plaa}, J., {Kaastra}, J.~S., {et~al.} 2017,
  \bibinfo{title}{{Radial metal abundance profiles in the intra-cluster medium
  of cool-core galaxy clusters, groups, and ellipticals},} \aap, 603, A80,
  \dodoi{10.1051/0004-6361/201630075}

\bibitem[{G. {Migliori} {et~al.}(2007){Migliori}, {Grandi}, {Palumbo},
  {Brunetti}, \& {Stanghellini}}]{Migliori2007}
{Migliori}, G., {Grandi}, P., {Palumbo}, G. G.~C., {Brunetti}, G., \&
  {Stanghellini}, C. 2007, \bibinfo{title}{{Radio Lobes of Pictor A: An X-Ray
  Spatially Resolved Study},} \apj, 668, 203, \dodoi{10.1086/520870}

\bibitem[{G. {Miley}(1980){Miley}}]{Miley1980}
{Miley}, G. 1980, \bibinfo{title}{{The structure of extended extragalactic
  radio sources},} \araa, 18, 165, \dodoi{10.1146/annurev.aa.18.090180.001121}

\bibitem[{D. {Nagai} {et~al.}(2007){Nagai}, {Kravtsov}, \&
  {Vikhlinin}}]{Nagai2007}
{Nagai}, D., {Kravtsov}, A.~V., \& {Vikhlinin}, A. 2007,
  \bibinfo{title}{{Effects of Galaxy Formation on Thermodynamics of the
  Intracluster Medium},} \apj, 668, 1, \dodoi{10.1086/521328}

\bibitem[{J. {Rasmussen} \& T.~J. {Ponman}(2007){Rasmussen} \&
  {Ponman}}]{Rasmussen2007}
{Rasmussen}, J., \& {Ponman}, T.~J. 2007, \bibinfo{title}{{Temperature and
  abundance profiles of hot gas in galaxy groups - I. Results and statistical
  analysis},} \mnras, 380, 1554, \dodoi{10.1111/j.1365-2966.2007.12191.x}

\bibitem[{J.~S. {Sanders}(2006){Sanders}}]{Sanders2006}
{Sanders}, J.~S. 2006, \bibinfo{title}{{Contour binning: a new technique for
  spatially resolved X-ray spectroscopy applied to Cassiopeia A},} \mnras, 371,
  829, \dodoi{10.1111/j.1365-2966.2006.10716.x}

\bibitem[{J.~S. {Santos} {et~al.}(2008){Santos}, {Rosati}, {Tozzi},
  {B{\"o}hringer}, {Ettori}, \& {Bignamini}}]{Santos2008}
{Santos}, J.~S., {Rosati}, P., {Tozzi}, P., {et~al.} 2008,
  \bibinfo{title}{{Searching for cool core clusters at high redshift},} \aap,
  483, 35, \dodoi{10.1051/0004-6361:20078815}

\bibitem[{C.~L. {Sarazin}(1988){Sarazin}}]{Sarazin1988}
{Sarazin}, C.~L. 1988, {X-ray emission from clusters of galaxies}

\bibitem[{E. {Scannapieco} {et~al.}(2005){Scannapieco}, {Silk}, \&
  {Bouwens}}]{Scannapieco05}
{Scannapieco}, E., {Silk}, J., \& {Bouwens}, R. 2005, \bibinfo{title}{{AGN
  Feedback Causes Downsizing},} \apjl, 635, L13, \dodoi{10.1086/499271}

\bibitem[{T.~W. {Shimwell} {et~al.}(2022){Shimwell}, {Hardcastle}, {Tasse},
  {Best}, {R{\"o}ttgering}, {Williams}, {Botteon}, {Drabent}, {Mechev},
  {Shulevski}, {van Weeren}, {Bester}, {Br{\"u}ggen}, {Brunetti}, {Callingham},
  {Chy{\.z}y}, {Conway}, {Dijkema}, {Duncan}, {de Gasperin}, {Hale},
  {Haverkorn}, {Hugo}, {Jackson}, {Mevius}, {Miley}, {Morabito}, {Morganti},
  {Offringa}, {Oonk}, {Rafferty}, {Sabater}, {Smith}, {Schwarz}, {Smirnov},
  {O'Sullivan}, {Vedantham}, {White}, {Albert}, {Alegre}, {Asabere}, {Bacon},
  {Bonafede}, {Bonnassieux}, {Brienza}, {Bilicki}, {Bonato}, {Calistro Rivera},
  {Cassano}, {Cochrane}, {Croston}, {Cuciti}, {Dallacasa}, {Danezi}, {Dettmar},
  {Di Gennaro}, {Edler}, {En{\ss}lin}, {Emig}, {Franzen},
  {Garc{\'\i}a-Vergara}, {Grange}, {G{\"u}rkan}, {Hajduk}, {Heald}, {Heesen},
  {Hoang}, {Hoeft}, {Horellou}, {Iacobelli}, {Jamrozy}, {Jeli{\'c}},
  {Kondapally}, {Kukreti}, {Kunert-Bajraszewska}, {Magliocchetti}, {Mahatma},
  {Ma{\l}ek}, {Mandal}, {Massaro}, {Meyer-Zhao}, {Mingo}, {Mostert}, {Nair},
  {Nakoneczny}, {Nikiel-Wroczy{\'n}ski}, {Orr{\'u}}, {Pajdosz-{\'S}mierciak},
  {Pasini}, {Prandoni}, {van Piggelen}, {Rajpurohit}, {Retana-Montenegro},
  {Riseley}, {Rowlinson}, {Saxena}, {Schrijvers}, {Sweijen}, {Siewert},
  {Timmerman}, {Vaccari}, {Vink}, {West}, {Wo{\l}owska}, {Zhang}, \&
  {Zheng}}]{Shimwell2022}
{Shimwell}, T.~W., {Hardcastle}, M.~J., {Tasse}, C., {et~al.} 2022,
  \bibinfo{title}{{The LOFAR Two-metre Sky Survey. V. Second data release},}
  \aap, 659, A1, \dodoi{10.1051/0004-6361/202142484}

\bibitem[{A. {Shulevski} {et~al.}(2017){Shulevski}, {Morganti}, {Harwood},
  {Barthel}, {Jamrozy}, {Brienza}, {Brunetti}, {R{\"o}ttgering}, {Murgia},
  {White}, {Croston}, \& {Br{\"u}ggen}}]{Shulevski2017}
{Shulevski}, A., {Morganti}, R., {Harwood}, J.~J., {et~al.} 2017,
  \bibinfo{title}{{Radiative age mapping of the remnant radio galaxy B2
  0924+30: the LOFAR perspective},} \aap, 600, A65,
  \dodoi{10.1051/0004-6361/201630008}

\bibitem[{J. {Siegel} {et~al.}(2025){Siegel}, {Bigwood}, {Amon}, {McCullough},
  {Yamamoto}, {McCarthy}, {Schaller}, {Schneider}, \& {Schaye}}]{Siegel2025}
{Siegel}, J., {Bigwood}, L., {Amon}, A., {et~al.} 2025, \bibinfo{title}{{The
  suppression of the matter power spectrum: strong feedback from X-ray gas mass
  fractions, kSZ effect profiles, and galaxy-galaxy lensing},} arXiv e-prints,
  arXiv:2512.02954, \dodoi{10.48550/arXiv.2512.02954}

\bibitem[{R.~K. {Smith} {et~al.}(2001){Smith}, {Brickhouse}, {Liedahl}, \&
  {Raymond}}]{Smith2001}
{Smith}, R.~K., {Brickhouse}, N.~S., {Liedahl}, D.~A., \& {Raymond}, J.~C.
  2001, \bibinfo{title}{{Collisional Plasma Models with APEC/APED:
  Emission-Line Diagnostics of Hydrogen-like and Helium-like Ions},} \apjl,
  556, L91, \dodoi{10.1086/322992}

\bibitem[{S.~L. {Snowden} {et~al.}(2008){Snowden}, {Mushotzky}, {Kuntz}, \&
  {Davis}}]{Snowden2008}
{Snowden}, S.~L., {Mushotzky}, R.~F., {Kuntz}, K.~D., \& {Davis}, D.~S. 2008,
  \bibinfo{title}{{A catalog of galaxy clusters observed by XMM-Newton},} \aap,
  478, 615, \dodoi{10.1051/0004-6361:20077930}

\bibitem[{N. {Soker}(2016){Soker}}]{Soker2016}
{Soker}, N. 2016, \bibinfo{title}{{The jet feedback mechanism (JFM) in stars,
  galaxies and clusters},} \nar, 75, 1, \dodoi{10.1016/j.newar.2016.08.002}

\bibitem[{G.~S.~C. {Stewart} {et~al.}(2025){Stewart}, {Shabala}, {Turner},
  {Yates-Jones}, {Krause}, {Wong}, {Power}, \& {Hardcastle}}]{Stewart2025}
{Stewart}, G. S.~C., {Shabala}, S.~S., {Turner}, R.~J., {et~al.} 2025,
  \bibinfo{title}{{CosmoDRAGoN II: Remnant radio galaxies in group and cluster
  environments},} \pasa, 42, e152, \dodoi{10.1017/pasa.2025.10113}

\bibitem[{M. {Sun}(2009){Sun}}]{Sun2009}
{Sun}, M. 2009, \bibinfo{title}{{Every BCG with a Strong Radio Agn has an X-Ray
  Cool Core: Is the Cool Core-Noncool Core Dichotomy Too Simple?},} \apj, 704,
  1586, \dodoi{10.1088/0004-637X/704/2/1586}

\bibitem[{M. {Sun}(2012){Sun}}]{Sun12}
{Sun}, M. 2012, \bibinfo{title}{{Hot gas in galaxy groups: recent
  observations},} New Journal of Physics, 14, 045004,
  \dodoi{10.1088/1367-2630/14/4/045004}

\bibitem[{M. {Sun} {et~al.}(2007){Sun}, {Jones}, {Forman}, {Vikhlinin},
  {Donahue}, \& {Voit}}]{Sun2007}
{Sun}, M., {Jones}, C., {Forman}, W., {et~al.} 2007, \bibinfo{title}{{X-Ray
  Thermal Coronae of Galaxies in Hot Clusters: Ubiquity of Embedded
  Mini-Cooling Cores},} \apj, 657, 197, \dodoi{10.1086/510895}

\bibitem[{M. {Sun} {et~al.}(2003){Sun}, {Jones}, {Murray}, {Allen}, {Fabian},
  \& {Edge}}]{Sun03}
{Sun}, M., {Jones}, C., {Murray}, S.~S., {et~al.} 2003,
  \bibinfo{title}{{Chandra Observations of the Galaxy Cluster A478: The
  Interaction of Hot Gas and Radio Plasma in the Core, and an Improved
  Determination of the Compton y-Parameter},} \apj, 587, 619,
  \dodoi{10.1086/368300}

\bibitem[{M. {Sun} {et~al.}(2009){Sun}, {Voit}, {Donahue}, {Jones}, {Forman},
  \& {Vikhlinin}}]{Sun09}
{Sun}, M., {Voit}, G.~M., {Donahue}, M., {et~al.} 2009,
  \bibinfo{title}{{Chandra Studies of the X-Ray Gas Properties of Galaxy
  Groups},} \apj, 693, 1142, \dodoi{10.1088/0004-637X/693/2/1142}

\bibitem[{J.~L. {Tinker}(2021){Tinker}}]{Tinker2021}
{Tinker}, J.~L. 2021, \bibinfo{title}{{A Self-Calibrating Halo-Based Group
  Finder: Application to SDSS},} \apj, 923, 154,
  \dodoi{10.3847/1538-4357/ac2aaa}

\bibitem[{R.~J. {Turner}(2018){Turner}}]{Turner2018}
{Turner}, R.~J. 2018, \bibinfo{title}{{Duty-cycle and energetics of remnant
  radio-loud AGN},} \mnras, 476, 2522, \dodoi{10.1093/mnras/sty433}

\bibitem[{F. {Ubertosi} {et~al.}(2023){Ubertosi}, {Gitti}, \&
  {Brighenti}}]{Ubertosi2023}
{Ubertosi}, F., {Gitti}, M., \& {Brighenti}, F. 2023, \bibinfo{title}{{Chasing
  ICM cooling and AGN feedback from the macro to the meso scales in the galaxy
  cluster ZwCl 235},} \aap, 670, A23, \dodoi{10.1051/0004-6361/202244023}

\bibitem[{G.~M. {Voit}(2005){Voit}}]{Voit05}
{Voit}, G.~M. 2005, \bibinfo{title}{{Tracing cosmic evolution with clusters of
  galaxies},} Reviews of Modern Physics, 77, 207,
  \dodoi{10.1103/RevModPhys.77.207}

\bibitem[{R. {Willingale} {et~al.}(2013){Willingale}, {Starling}, {Beardmore},
  {Tanvir}, \& {O'Brien}}]{Willingale2013}
{Willingale}, R., {Starling}, R.~L.~C., {Beardmore}, A.~P., {Tanvir}, N.~R., \&
  {O'Brien}, P.~T. 2013, \bibinfo{title}{{Calibration of X-ray absorption in
  our Galaxy},} \mnras, 431, 394, \dodoi{10.1093/mnras/stt175}

\end{thebibliography}
\bibliographystyle{aasjournalv7}



\end{document}